\documentclass[aps,prb,twocolumn]{revtex4}
\usepackage{graphicx}
\usepackage{epsfig}
\newcommand{\bra}[1]{\langle {#1} |}
\newcommand{\ket}[1]{| {#1} \rangle}

\begin{document}
\bibliographystyle{revtex}
\title{ELECTRON-PHONON INTERACTION IN ULTRASMALL-RADIUS CARBON NANOTUBES}
\author{Ryan Barnett, Eugene Demler, and Efthimios Kaxiras}
\affiliation{Department of Physics, Harvard University, Cambridge MA 02138}
                \date{\today}

\begin{abstract}
We perform analysis of the band structure, phonon dispersion, and
electron-phonon interactions in three types of small-radius
carbon nanotubes. 
We find that the (5,5) can be described well by the zone-folding
method and the  
electron-phonon interaction is too small to support either
a charge-density wave or superconductivity at realistic temperatures.
For ultra-small (5,0) and (6,0) nanotubes we find that the
large curvature makes these tubes metallic with a large density of
states at the Fermi energy and leads to unusual electron-phonon
interactions, with the dominant coupling coming from the out-of-plane
phonon modes.  By combining the frozen-phonon approximation with the
RPA analysis of the giant Kohn anomaly in 1d we find parameters of the
effective  Fr\"{o}lich Hamiltonian  for the conduction electrons.  
Neglecting Coulomb interactions, we find that the (5,5) CNT
remains stable to instabilities of the Fermi surface down to very
low temperatures while for the (5,0) and (6,0) CNTs a CDW instability
will occur.  
When we
include a realistic model of Coulomb interaction we find that the
charge-density wave remains dominant in the (6,0) CNT with
$T_{\rm CDW}$ around 5 K while the 
charge-density wave instability is suppressed to very low temperatures in 
the (5,0) CNT, making 
superconductivity dominant with transition temperature around one
Kelvin.

\end{abstract}

\pacs{PACS numbers:}

\maketitle

\section{INTRODUCTION}
\label{intro}

It has been over a decade since the discovery of carbon nanotubes
(CNTs) \cite{Ijima91} and the interest level in these systems
continues to be high.  The majority of theoretical
work on CNTs focuses on
understanding the effects of the electron-electron interactions using
the celebrated Luttinger liquid theory.\cite{Egger00}  Experimental
observation of superconductivity in ropes of nanotubes  \cite{Kociak01}  and
small-radius nanotubes in a zeolite matrix \cite{Tang01} has also
motivated theoretical studies of the electron-phonon interactions
(EPI), including the analysis of charge density wave (CDW)
\cite{Mintmire92,Huang96,Sedeki00,Dubay02}
and superconducting (SC)
\cite{Benedict95,Sedeki02,Byczuk02,Gonzalez02,Kamide03}
instabilities.  In this work we study the electron-phonon interactions 
in CNTs and discuss possible instabilities to the CDW
and SC orders.  Our  approach provides 
reliable parameters for the effective Hamiltonians we use
in contrast to the Luttinger liquid treatments
where obtaining such accurate quantities is quite difficult.

A conventional starting point
for discussing the electron-phonon
interaction in solids
is the Fr\"{o}\-lich Hamiltonian \cite{Schrieffer64}
\begin{eqnarray}
\label{frolich}
{\cal H}&=&\sum_{k\tau \sigma}\varepsilon_{k \tau}
c_{k\tau\sigma}^{\dagger}c_{k\tau\sigma}+
\sum_{q\mu}\Omega^0_{q\mu}(a_{q\mu}^{\dagger}a_{q\mu}+
\frac{1}{2})
\nonumber
\\
&+&\sum_{k\tau k'\tau' \sigma\mu}g_{k \tau k'\tau' \mu}
c_{k\tau\sigma}^{\dagger}c_{k'\tau'\sigma}(a_{q\mu}+a_{-q\mu}^{\dagger}).
\end{eqnarray}
Here $c^\dagger_{k\tau \sigma}$  creates an electron
with quasimomentum $k$ in band $\tau$ with spin $\sigma$,
$a^\dagger_{q\mu}$ creates a phonon
with lattice momentum $q$ and polarization $\mu$,
and $q=k-k'$ modulo a reciprocal lattice vector.
The energies of electron quasiparticles and phonons (in the absence
of EPC) are
given by $\varepsilon_{k\tau}$ and $\Omega^0_{q\mu}$ respectively.
The EPC vertex is given by
\begin{equation}
\label{vertex}
g_{k\tau k'\tau'\mu}=\sqrt{\frac{1}{2\Omega^0_q M N N_c}}
M_{k\tau k'\tau'\mu}
\end{equation}
with
\begin{equation}
\label{eq:Mkk}
M_{k\tau k'\tau'\mu}
=N \bra{\psi_{k\tau}}
\sum_{i}\frac{\partial V}{\partial {\bf R}_{0i}} \cdot
\hat{\epsilon}_{q\mu}(i)\ket{\psi_{k'\tau'}}.
\end{equation}
Here $\ket{\psi_{k\tau}}=c^{\dagger}_{k\tau}\ket{0}$ is a
quasistationary electron state
in band $\tau$ with quasimomentum $k$,
$\hat{\epsilon}_{q\mu}(i)$ is the phonon
polarization vector on atom $i$ in the
unit cell, $N_c$
is the number of atoms per unit cell, $M$
is the mass of a single C atom, $N$ is the total
number of unit cells in the system, and $\frac{\delta V}{\delta {\bf R}_{0i}}$
is the derivative of the crystal potential with respect
to the ion position ${\bf R}_{0i}$.

A common approach to obtaining parameters of the Hamiltonian Eq.
(\ref{frolich}) for the CNTs is the zone-folding method (ZFM).
\cite{Saito98}  The essence of this method is to take the electron band
structure and the phonon dispersion for graphene and quantize
momenta in the direction of the  wrapping.  The main results
of such a procedure may be summarized as follows.  The only
bands crossing the Fermi level in graphene are the bonding and the
antibonding combinations of the atomic $p_z$ orbitals. Hence, the
zone-folding method predicts that these are the only bands which may cross
the Fermi level in carbon nanotubes. The condition for the
quantized momenta to cross the Dirac points of the graphene gives the
condition for the (N,M) CNT to be metallic: $N-M$ should be
divisible by 3.  The ZFM also predicts that the
electron-phonon coupling in the CNTs should be
dominated by the in-plane optical modes. This follows from the fact
that the latter have the largest effect on the overlaps between the
$p_z$ orbitals of the neighboring carbon atoms.

\begin{figure}
\includegraphics[width=3in]{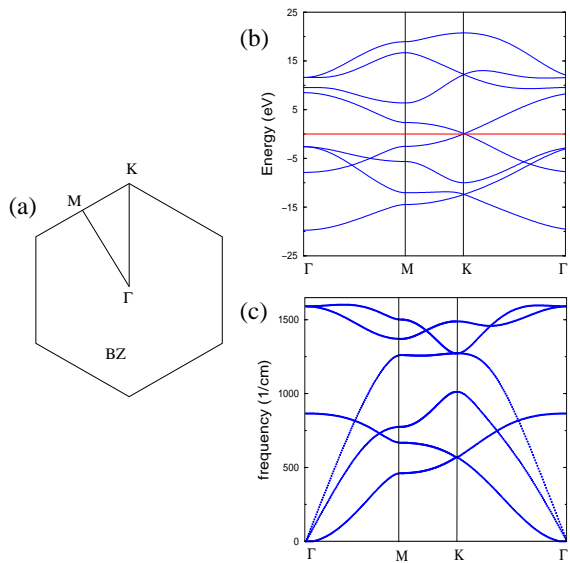}
\caption{The first Brillouin zone (a), electronic band
structure (b), and phonon dispersion (c) of graphene.}
\label{fig:graphite_pho_bs}
\end{figure}

While the ZFM was shown to provide a quantitatively accurate
description of the larger radius nanotubes, it is expected to
fail as the radius of the nanotubes is decreased and the curvature of
the C-C bonds becomes important. Determining the band structure, the
phonon dispersion, and the electron-phonon coupling of the small
radius CNTs requires detailed microscopic calculations.  In this paper
we use the empirical tight-binding model \cite{Mehl96}
to provide such analysis for
three types of small-radius nanotubes: (5,0) with the diameter 3.9
\AA, (6,0) with the diameter 4.7 \AA, and (5,5) with the diameter
6.8 \AA.  We find that the large curvature of the C-C bonds leads
to qualitative changes in the band structure of the (5,0) and (6,0)
nanotubes (band structures of the (6,0) and (7,0) nanotubes have been
discussed previously in \cite{Blase94}).  For example, the (5,0) CNT becomes
metallic from strong hybridization between the $\sigma$ and $\pi$ bands
(see Fig.~\ref{fig:5,0comb}).  Frequencies of the phonon modes in small
radius CNTs are also strongly renormalized from their values in
graphene. Not only does the out-of-plane acoustic mode become a finite
frequency breathing mode, \cite{Saito98} but even the optical modes change
their energy appreciably (see e.g.  Fig.~\ref{fig:pho5,0}). Finally,
the electron-phonon coupling changes qualitatively in the small-radius
CNTs. It is no longer dominated by the in-plane optical modes but by
the out-of-plane optical modes which oscillate between the $sp_2$
bonding of graphene and the $sp_3$ bonding of diamond (see discussion
in Sec.~\ref{sec:discussion}).
We find that the strong effects of the
CNT curvature decrease rapidly with increasing the tube
radius. Already for the (5,5) nanotubes the ZFM gives a
fairly accurate description of the band structure as well as the
electron-phonon interactions.

Determining parameters of the Fr\"{o}lich Hamiltonian for a one-dimensional
system is not as straightfoward as for two and three-dimensional
metals.  Traditional methods for analyzing EPI
from first-principles calculations are mean-field and, therefore,
suffer from instabilities intrinsic to one-dimensional systems. In
particular, the frozen-phonon approximation, which is commonly used to
determine the phonon frequencies, $\Omega^0_{q\mu}$, in Eq.~(\ref{frolich})
gives imaginary frequencies close to the nesting wave vector
$q=2k_F$. This is the result of the giant Kohn anomaly, \cite{Kohn59} which
corresponds to the Peierls instability of the one-dimensional
electron-phonon system.\cite{Peierls55}  An important result of our paper is
that we developed a new formalism, which combines the frozen-phonon
approximation with the  Random-Phase
Approximation (RPA) analysis of the EPI. This allows us to extract
effective non-singular parameters of the Fr\"{o}lich Hamiltonian from
first-principles calculations or from the empirical tight-binding
model.  This technique should be applicable to many systems other than
carbon nanotubes.

After determining parameters of the Fr\"{o}lich Hamiltonian
Eq.~(\ref{frolich}) for the (5,0), (6,0), and (5,5) CNTs we discuss
possible superconducting and charge-density wave instabilities in
these systems. We find that neglecting the residual Coulomb
interaction leads to much stronger CDW instabilities in all three cases
(in such analysis Coulomb interaction is included only at the
mean-field level via the energy of the single-particle
quasi-stationary states, $\varepsilon_{k \tau}$). In the mean-field
approximation we find the onset of the Peierls instability at
temperatures 160, 5, and $10^{-14}$ K  for (5,0), (6,0), and (5,5)
CNTs respectively. However, including the
Coulomb interactions at the RPA level \cite{Levin74} can lead to a 
stronger suppression of the CDW transition temperatures, $T_{\rm CDW}$, than
the superconducting $T_{\rm SC}$.  For instance, we find by
using the model Coulomb interaction of Ref.~\onlinecite{Egger98} 
that for the (5,0) CNT, the CDW transition is 
suppressed to very low temperatures
 while superconductivity becomes the dominant phase
with transition temperature of $T_{\rm SC} \approx 1$ K.

This paper is organized as follows.  In Sec.~\ref{sec:extract} we
discuss our method for extracting parameters for the one-dimensional
Fr\"{o}lich Hamiltonian.  We then apply this method to the (5,0), (6,0), 
and (5,5) CNTs in Sec.~\ref{sec:results}.  In Sec.~\ref{sec:instabilities},
we use the constructed Hamiltonian for these CNTs to study their
instabilities toward superconductivity and charge-density wave
states.  The effect of introducing the residual Coulomb interacting
between electrons is covered in Sec.~\ref{sec:coulomb}.  Finally
all of the results are discussed and summarized in 
Sec.~\ref{sec:discussion}.

\section{EXTRACTING PARAMETERS OF THE EFFECTIVE FR\"{O}LICH
HAMILTONIAN FROM THE FIRST PRINCIPLE CALCULATIONS}
\label{sec:extract}

Now we discuss our methods for calculating input parameters to the
Fr\"{o}lich Hamiltonian Eq.~(\ref{frolich}) for the representative nanotubes.
Our analysis relies on the the empirical tight-binding model
\cite{Mehl96} but it is easily amenable to any density-functional
theory \cite{Hohenberg64, Kohn65} treatment of the system.

\subsection{Band Structure}

To compute the electronic structure of the CNTs we study, we use the
NRL tight-binding method \cite{Mehl96} which has been tested and
provided accurate results on a variety of materials.  In this method,
the Slater-Koster tight-binding matrix elements are parametrized and
are fit to reproduce the first-principles density-functional band
structures and total energies, with around 70 adjustable parameters
per element.

We study the (5,0), (6,0), and (5,5) CNTs which are shown in
Figs.~\ref{fig:5,0comb},\ref{fig:6,0comb}, and \ref{fig:5,5comb}.
  The smallest possible unit cells for these
CNTs contain 20, 24, and 20 atoms respectively.  These CNTs are
relaxed by minimizing their total energy per unit cell with respect to
the atomic coordinates using 35 k-points in the first Brillouin zone.
Matrix elements between neighboring atoms of up to 5.5~\AA~were used.
The calculations were performed on an orthorhombic lattice with spacing
between parallel CNTs of 16~\AA, a distance sufficiently large to
ensure negligible dispersion from inter-tube hopping.  Once the CNTs
are relaxed, the band structure is calculated.

\subsection{The phonon modes}

To calculate the electron-phonon coupling vertices and the phonon 
frequencies which will be discussed in the subsequent sections, 
one needs to have the ionic displacements corresponding
to the normal vibrational modes of the CNT.  As pointed out previously,
\cite{Sanchez-Portal99, Saito98} we find that it is typically
sufficient to use the zone-folded modes of a graphene sheet, even
for the small-radius CNTs we study as will be discussed below.

\begin{figure}
\includegraphics[width=3in]{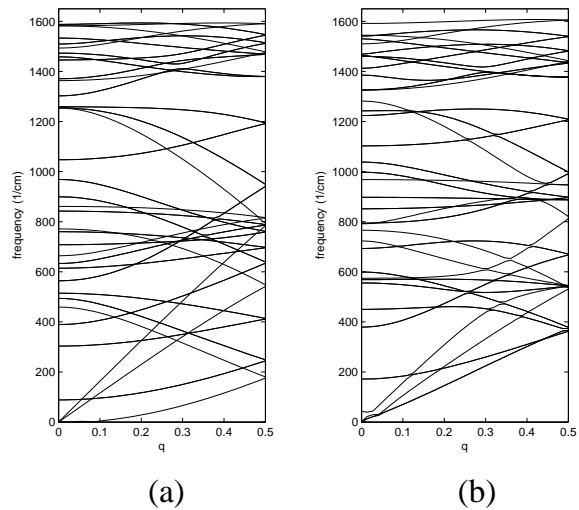}
\caption{The phonon dispersions of a (5,0) CNT determined
by (a) the zone-folding method and (b) diagonalizing the full
dynamical matrix of the CNT.}
\label{fig:modes_fold}
\end{figure}

Following the method used in the book of Saito \emph{et al.},
\cite{Saito98} we have computed the $60 \times 60$ dynamical
matrix of a (5,0) CNT and in Fig.~\ref{fig:modes_fold} we
compare the resulting phonon dispersions with the zone-folding
results.  The ionic displacement modes obtained by the two different methods
are very similar except for a few special cases.  
For instance, the zone-folding results give three acoustic modes which
correspond to translating the graphene sheet in 
different directions.  Upon rolling the graphene sheet, these
modes get mapped to two acoustic modes corresponding to
rotation about the CNT axis and translation along the
CNT axis and the optical breathing mode.
Conversely, diagonalizing the dynamical matrix of the
CNT gives four acoustic modes corresponding to translations
in three directions and the rotating mode (actually using
the method of Ref.~\onlinecite{Saito98}, one obtains a small spurious
frequency for the rotating mode as pointed in this reference).  
Upon unrolling
the CNT to the graphene sheet, the rotating mode and the
mode corresponding to translation along the CNT axis will
become acoustic modes of the graphene sheet.  However,
the two CNT translational modes which are perpendicular
to the CNT axis will get mapped to ionic displacements which
are not eigenmodes of the graphene sheet which are 
mixtures of in-plane and out-of-plane oscillations.  
In addition, using the dynamical matrix of the CNT, we
find that there is mixing between the breathing and 
strething modes around $k=0.3$.  In this vicinity, there
is level repulsion from the lifting of the degeneracy of these
modes.  Away from this point, the modes are, to a good approximation,
decoupled.

In our analysis of the electron-phonon coupling we use the
displacements obtained from the zone-folding method to 
simplify the calculations, as well as to give a clear conceptual
picture.  We then check that none
of the important electron-phonon couplings come from 
any of the few graphene modes for which the zone-folding
method breaks down.

\subsection{The electron-phonon coupling vertices}

The electron-phonon coupling (EPC) matrix
in Eq.~(\ref{eq:Mkk}) can be evaluated by
using the finite difference formula
\begin{equation}
\label{Mkk'}
M_{k\tau k'\tau'\mu}
=\frac{1}{u}\bra{\psi_{k\tau}}(V_{q\mu}-V_{0})\ket{\psi_{k'\tau'}}
\end{equation}
where $V_{q\mu}$ and $V_{0}$ are the perturbed and the
unperturbed lattice potentials respectively.
A method for calculating the expression (\ref{Mkk'})
with a plane-wave basis set
was previously developed.\cite{Lam86} In this paper
we extend this procedure to tight-binding models.
We introduce the standard tight-binding notation
\begin{equation}
\ket{\psi_{k\tau}}=\sum_{il}A_{k\tau il}\ket{\chi_{kil}}.
\end{equation}

\begin{equation}
\ket{\chi_{kil}}=\frac{1}{\sqrt{N}}\sum_{n}e^{ik\cdot R_{n}}\ket{\phi_{nil}}.
\end{equation}
Here $\ket{\phi_{ni}}$ are the electron states for isolated carbon
atoms, $n$ runs over unit cells, $i$ runs over basis vectors in
the unit cell, and $l$ runs over orbital type. We find (for details, see
Appendix \ref{Appendix:Mkk'})
\begin{equation}
M_{k\tau k'\tau'\mu}=\frac{1}{u}\sum_{ili'l'}A^{*}_{k\tau il}\bra{\chi_{kil}^{q\mu}}
({\cal H}^{q\mu}-E_{F})\ket{\chi_{k'i'l'}^{q\mu}}A_{k'\tau' i'l'}.
\end{equation}
This expression can be computed by evaluating the tight-binding
Hamiltonian and overlap matrices for the distorted lattice, evaluating
the coefficients $A_{ki}$ and $A_{k'i'}$ of the wave functions for the
undistorted lattice, and performing the above sum.

In all the calculations presented in this paper we used the ZFM to
find phonon eigenvectors in the nanotubes starting from the phonon
eigenvectors in graphene.\cite{Saito98}  The latter have been
obtained using the $6 \times 6$ dynamical matrix of graphene given
in Ref.~\onlinecite{Jishi93}.  We emphasize that we use the
ZFM only to find the
phonon eigenvectors in small nanotubes, but not the phonon
frequencies. The frequencies are affected strongly by the CNT
curvature, and should be computed directly. This is
discussed in detail in Sec.~\ref{sec:frequencies} and
Sec.~\ref{sec:results}.

\subsection{Phonon frequencies}
\label{sec:frequencies}

A standard method of calculating the bare phonon frequencies
$\Omega_{q\mu}^{0}$ in Eq.~(\ref{frolich}) is the frozen-phonon approximation
 (FPA).\cite{Yin82} In this approach
\begin{equation}
\label{Eq:FPA}
\Omega_{q\mu}= \frac{1}{u\sqrt{M N_c}}\sqrt{\Delta
E_{\cos}(q) + \Delta E_{\sin}(q)}
\end{equation}
where $u$ is the amplitude of the displacement, and $\Delta E_{\cos}(q)$
and $\Delta E_{\sin}(q)$ are the energy differences per unit cell between
the distorted and equilibrium lattice structures where the distortion
corresponds to the real and imaginary parts of $\delta{\bf R}_{ni}=u
e^{iq R_{n}} \hat{\epsilon}_{q\mu}(i)$ respectively.  When we apply
this procedure to one-dimensional CNTs, we find that
$\Delta E_{\cos}(q)+\Delta E_{\sin}(q)$ becomes
negative around certain wave vectors (see
e.g.~Fig.~\ref{fig:pho5,0}). A closer inspection shows that such
anomalous softening always corresponds to one of the $2 k_F$
wave vectors of the electron bands indicating the presence of
the giant Kohn anomaly.

It is important to realize that the divergence
of $\Omega_{q\mu}$ obtained in the FPA does not imply
the divergence of $\Omega^0_{q\mu}$ in the
Fr\"{o}lich Hamiltonian Eq.~(\ref{frolich}).
The frequencies $\Omega_{q\mu}$ are calculated
{\it after} the electron-phonon
interaction in Eq.~(\ref{frolich})
have been included, which gives
anomalous softening at $2 k_F$
due to the well-known
Peierls instability of electron-phonon
systems in 1d. In two and three dimensional
systems renormalization of the phonon frequency
by electrons in the conduction band is typically negligible.
So, one can use phonon energies obtained in the FPA
as a direct input into the Fr\"{o}lich Hamiltonian.
By contrast, nesting of the one-dimensional
Fermi surfaces, leads to dramatic renormalization
of the phonon dispersion by electrons in the conduction
band.

\begin{figure}
\includegraphics[width=3in]{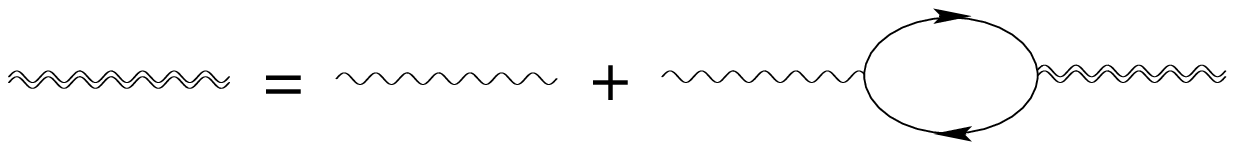}
\caption{The phonon propagator evaluated within the RPA.}
\label{fig:phoRPA}
\end{figure}

To extract the bare phonon frequency
$\Omega^0_{q\mu}$ from the numerically computed
$\Omega_{q\mu}$, we point out a connection
between the FPA
and the RPA
for the   Fr\"{o}lich Hamiltonian. For negligible interband coupling
(this condition is satisfied for all modes
showing the giant Kohn anomaly, which we discuss in this paper)
Dyson's equation for the phonon propagator $D(q,i\nu_m)$, as shown
in \ref{fig:phoRPA} is given by
\begin{equation}
\label{eq:Dyson}
D_\mu(q,i \nu_{m})=D_{0\mu}(q,i \nu_{m})\left(1+
\Pi_\mu(q,i \nu_{m})D_\mu(q,i \nu_{m})\right).
\end{equation}
Here $\nu_m=2\pi m T$  are the bosonic Matsubara frequencies
and
\begin{equation}
\label{eq:D0}
D_{0\mu}(q,i \nu_{m})=
\frac{2\Omega_{q\mu}}{(i\nu_m)^{2}-(\Omega_{q\mu}^{0})^2}.
\end{equation}
is the non-interacting phonon Green's function.
The phonon self-energy evaluated in the RPA  is given by
\begin{equation}
\label{polarization}
\Pi_\mu(q,i\nu_m)=
2  T \sum_{n p \tau} |g_{p\tau p+q \tau \mu}|^{2}
G_{0\tau}(p+q,i\omega_{m+n})G_{0\tau}(p,i \omega_n).
\end{equation}
where non-interacting electronic Green's functions are given by
$
G_{0\tau}(p,i\omega_n)=(i\omega_n-\varepsilon_{p\tau})^{-1}
$
and $\omega_n=\pi(2n+1)T$ for integer $n$ are
the fermionic Matsubara frequencies.  Summing
over $n$, we obtain for Eq.~(\ref{polarization})
\begin{equation}
\Pi_\mu(q,i\nu_m)=2\sum_\tau |g_{q\tau\mu}|^{2}\chi_{0\tau}(q,i\nu_m)
\end{equation}
where the bare susceptibility is given by
\begin{equation}
\label{susc}
\chi_{0\tau}(q,i\nu_m)=\sum_{p}\frac{f(\varepsilon_{p\tau})-f(\varepsilon_{p+q\tau})}
{i\nu_m+\varepsilon_{p\tau}-\varepsilon_{p+q\tau}}.
\end{equation}
with $f(\varepsilon_{p\tau})=(1+e^{\beta\varepsilon_{p\tau}})^{-1}$
being the Fermi-Dirac distribution
function.

The poles of the phonon Green's function $\Omega_{q\mu}$ (we
put $i\nu_m\rightarrow \Omega_{q\mu}$ in $D_\mu(q,i\nu_m)$),
which give the dressed
phonon frequencies, will satisfy the equation
\begin{equation}
\label{poles}
\left( \Omega_{q\mu}\right)_{\rm RPA}^{2}=\left( \Omega_{q\mu}^{0}\right)^{2}
+2\Omega^0_{q\mu}\Pi_\mu(q,\Omega_{q\mu}).
\end{equation}
Due to the large energy difference between electrons and phonons,
it is typically a good approximation to set $\Omega_{q\mu}\rightarrow 0$
in $\Pi_\mu(q,\Omega_{q\mu})$.
This approximation results in an expression
that can be derived by doing stationary second-order
perturbation theory to obtain the change in energy
due to the presence of the phonon.  That is, setting $\Omega_{q\mu}\rightarrow 0$
in $\Pi(q,\Omega_{q\mu})$  corresponds to the
frozen-phonon approximation
\begin{equation}
\label{Eq:polesFPA}
\left( \Omega_{q\mu}\right)_{\rm FPA}^{2}
=\left( \Omega_{q\mu}^{0}\right)^{2}
+2\Omega^0_{q\mu}\Pi_\mu(q,0).
\end{equation}
We can typically approximate well the quasiparticle energy
by a plane-wave state with given effective mass $m^{*}$.  Then,
by incorporating the FPA, at
zero temperature the integral in Eq.~(\ref{susc}) can be done which
will enable us to obtain
\begin{eqnarray}
\label{fitCDW}
&&\left(\Omega_{q\mu}\right)_{\rm FPA}^{2} =
\left(\Omega_{q\mu}^{0}\right)^{2}
\nonumber\\
&+&\sum_\tau |M_{2k_{\rm F}\tau\mu}|^{2}
\frac{2 m^{*}a}{\pi  M N_{c}k_{{\rm F}\tau}}
\log \left|\frac{2k_{{\rm F}\tau}-
q}{2 k_{{\rm F}\tau}+q} \right|
\end{eqnarray}
This expression explicitly shows the logarithmic divergences in the
phonon dispersion at the nesting wave vectors of the Fermi
surface. This is the famous Peierls instability to a CDW state.
Our procedure for  determining the elusive
undressed frequencies is then as follows. We take
$\Omega_{q\mu}$ obtained from the FPA and fit them with the
expression Eq.~(\ref{fitCDW}) using $\Omega_{q\mu}^{0}$ as an
adjustable parameter.
The coefficients of the log divergences at the nesting wave vectors of
the Fermi surface are fixed by the effective masses $m_{\tau}^{*}$
and $k_{{\rm F}\tau}$ (known from the band structure) and the computed
EPC matrix elements $M_{2k_{F}\tau\mu}$.  In all cases we found excellent
agreement of the calculated FPA frequencies
with Eq.~(\ref{fitCDW}) in the vicinity of the singular
points, which provides a good self-consistency check for our analysis.

\section{RESULTS FOR REPRESENTATIVE NANOTUBES}

\label{sec:results}

\subsection{(5,0) nanotube}

The zig-zag (5,0) CNT has a diameter of around 3.9 \AA~making it
close to the theoretical limit.\cite{Peng00}  Nanotubes of
this size have been experimentally realized through growth
in the channels of a zeolite host.\cite{Tang01}
Through the Raman measurement of the frequency
of the radial breathing mode, the (5,0) CNT is thought to be
a likely candidate structure for these experiments.\cite{Li01}

\begin{figure}
\includegraphics[width=3in]{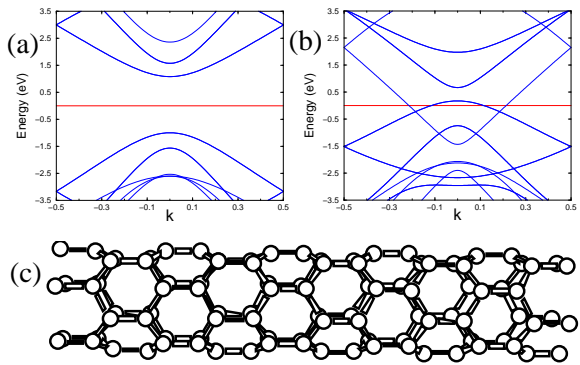}
\caption{The band structure of the (5,0) CNT obtained through
zone-folding (a) and calculated directly (b) along with the
atomic structure (c).}
\label{fig:5,0comb}
\end{figure}

We first compute the band structure of this tube by using the
zone-folding method.\cite{Saito98}  To do this, we use the
band structure of graphene, which is shown in Fig.~\ref{fig:graphite_pho_bs},
computed by using the NRL tight
binding method.  Shown
in this figure are four valence bands and four conduction bands,
coming from the three $sp_2$ and one $p_z$ bonding and antibonding
states respectively.  There is a degeneracy between the $p_z$
bonding and antibonding states at the Fermi energy at the $K$ point in
the first Brillouin zone which accounts for the semimetallic behavior
of graphene.  The zone-folding band structure of the (5,0) CNT
is shown in the right of Fig.~\ref{fig:5,0comb}.  Since $5/3$
is not an,
zone-folding predicts this CNT to be semiconducting.

Fig.~\ref{fig:5,0comb} (b) shows the band structure of the (5,0)
CNT calculated directly by using a unit cell of 20 atoms.  One sees
that there are significant qualitative differences between the
two band structures, one being that the directly computed band structure
predicts metallic behavior.  The inner band (with smaller Fermi point
$k_{F}^{\rm A}$) is doubly degenerate while the outer band (with larger Fermi
point $k_{F}^{\rm B}$) is nondegenerate.  The strong curvature effects
causes hybridization between $\sigma$ and $\pi$ bands, pushing them through the
Fermi energy and therefore making the tube metallic.  Furthermore,
for the (5,0) CNT, we see that inner band is close to the Van
Hove singularity at $k=0$, which produces a large density of states
at the Fermi energy.  The calculated density of $\nu(0)=0.16$ states/ eV /
C atom is around a factor of five larger than that of larger radius
metallic armchair CNTs.

After the band structure is calculated,
we consider all possible scattering processes of electrons
between Fermi points
$-k_F^{\rm B},-k_F^{\rm A},k_F^{\rm A}$, and $k_F^{\rm B}$
due to phonons with wave vectors $q$ that satisfy the momentum conservation
condition.
As a starting point
for the phonon spectrum, we use the dynamical matrix of Jishi \emph{et al.} \cite{Jishi93}
which uses a fourth nearest-neighbor model,
and we employ the zone-folding method.  The reproduced phonon dispersion
of graphene is shown in Fig.~\ref{fig:graphite_pho_bs}.
For a given process,
we calculate the coupling for all of the $3 \times N_c$ distinct
phonon modes where $N_c=20$ is the number of atoms per unit cell.
Shown in Fig.~\ref{fig:coup5,0} is an example of the outcome for one of these
calculations.  Shown is the coupling for the outer band $2 k_F^{\rm A}$
processes vs. graphene frequency.  One can immediately see that most
couplings vanish which can be explained by symmetry of the
electronic wave functions and the phonon modes.

\begin{figure}
\includegraphics[width=3in]{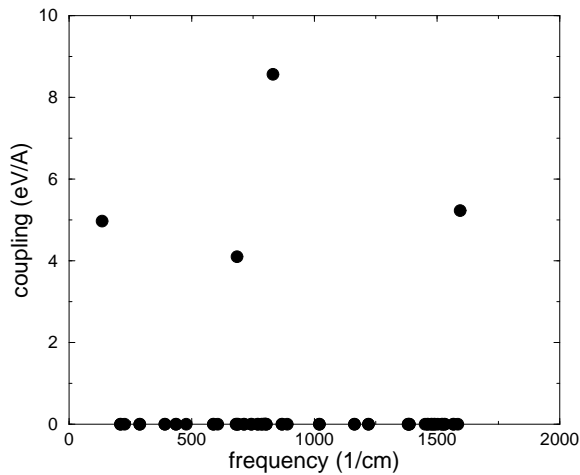}
\caption{The coupling $M_{kk'}$ for the outer band $2k_F^{\rm B}$ process
for each of the 3 $\times$ 20 = 60 phonon modes respectively vs.
graphene frequency.}
\label{fig:coup5,0}
\end{figure}

\begin{figure}
\includegraphics[width=3in]{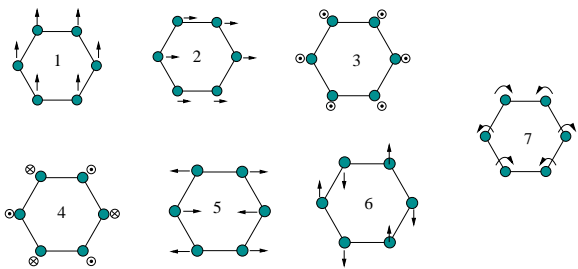}
\caption{1-6: The phonon modes at the $\Gamma$ point in the first Brillouin
 zone of graphene.
7:  An in-plane optical phonon mode at the $K$ point of the first
Brillouin zone of graphene.  The out-of-plane optical mode 4 is the leading
cause of the CDW instability in the (5,0) and (6,0) CNTs.}
\label{fig:modes}
\end{figure}

To keep this paper concise, we cannot present all of the coupling
results for each scattering process.  Instead, we show the
most dominant couplings.
These dominant couplings were found to be from
intraband $2k_F$ processes.  The largest couplings for the (5,0)
CNT occur for phonons along the $\Gamma M$ line of graphene
at the appropriate wave vector corresponding to the particular
$2k_F$.  For the inner band,
the largest couplings, in descending order, occur for the out-of-plane
optical mode, the radial breathing mode, and the in-plane acoustic
stretching mode.
For the outer (with larger $k_F$) band, the dominant couplings occur
for the out-of-plane optical, an in-plane optical, the radial breathing,
and in-plane stretching modes.
These results are summarized in Fig.~\ref{fig:modes}
and Table \ref{table:modes5,0}.
Although the magnitunde of the dominant
coupling matrix element for the outer band is larger than that of the inner
band, the inner band processes are significantly more important
in the study of instabilities because their contribution to the total
density of states at the Fermi energy is significantly larger than
that of the outer band.  This is due to the small Fermi velocity
of the inner band and its degeneracy.

\begin{table}
\begin{tabular}{c|c|c|c|}
(5,0) & mode & $\omega_{q}^{graph}$  (cm$^{-1}$)& $M_{kk'}$(eV/\AA)\\ \hline
$2k_{F}^{\rm A}$&4&853&5.55\\
&3&39&4.46\\
&5&1588&4.24\\ \hline
$2k_{F}^{\rm B}$&4&829&8.56\\
&5&1593&5.23\\
&3&133&4.97\\
&2&684&4.10\\
\end{tabular}
\caption{Calculated values for the dominant coupling processes
for the (5,0) CNT.  The numbering scheme here
corresponds to that given in Fig.~\ref{fig:modes}.
$2k_{F}^{\rm A}$ and $2k_{F}^{\rm B}$ correspond to inner and
outer band processes respectively.  Phonon frequencies are given for
graphene.}
\label{table:modes5,0}
\end{table}

It is interesting to note
that the phonons that have the strongest coupling to electrons
at the Fermi surface are out-of-plane modes.  This is different than
intercalated graphene where in-plane phonon modes are responsible
for superconductivity.\cite{Dresselhaus81}  The fact that the out-of-plane
modes are the most important for this CNT are presumably due to the
large curvature effects.  For instance, we find that the bond angles
of the relaxed (5,0) CNT structure (having the values
of $119.4^\circ$ and $111.9^\circ$)are intermediate between the
$sp_2$ bond angle (found in graphene) of $120^\circ$ and the $sp_3$ bond angles
(found in diamond) of $109.4^{\circ}$.

Now we calculate the CNT phonon frequencies by using the frozen-phonon
approximation with the eigenvectors from graphene.  The circles
shown in Fig.~\ref{fig:pho5,0} are the frequencies obtained for phonon
modes along the $\Gamma M$ line of graphene for the out-of-plane
optical mode which was found to be the most important mode.
First, we see that the calculated FPA frequencies are significantly
lower than the corresponding ones in graphene.  This can be understood
as follows.  The strong curvature of the nanotube changes the C-C
bonds so that they are in an intermediate regime between the
$sp_2$ bonding (found in graphene) and $sp_3$ bonding (found
in diamond).  The out-of-plane optical mode oscillates between these
two bonding configurations and is therefore significantly softened.
Next, we notice that there are divergences at $q=2k_F^{\rm A}$ and
$q=2k_F^{\rm B}$.  This result is the giant Kohn anomaly.

\begin{figure}
\includegraphics[width=3in]{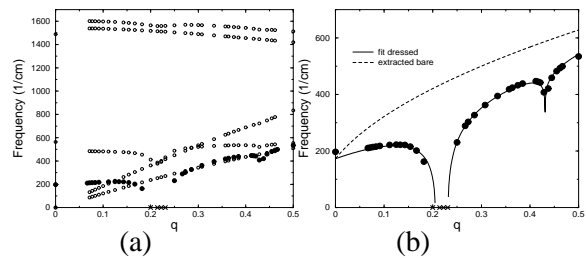}
\caption{(a): Phonon dispersion for the (5,0) CNT along the
$\Gamma M$ line of graphene.  The X's denote values
for which the frozen-phonon approximation gave imaginary frequencies
for the out-of-plane optical mode in the vicinity of $2k_F^{\rm A}$.
(b):  The mode showing the most softening fit to the RPA
expression.}
\label{fig:pho5,0}
\end{figure}

To extract the bare phonon frequency of the Fr\"{o}lich
Hamiltonian Eq.~\ref{frolich} for the (5,0) CNT we follow
the procedure discussed in Sec.~\ref{sec:frequencies}.
The dressed
phonon frequencies are given by
\begin{eqnarray}
\label{Eq:fit}
\left(\Omega_{q\mu}\right)^{2}=\left(\Omega_{q\mu}^{0}\right)^{2}&+&
D_{\rm A}
\log \left|\frac{2k_{F}^{\rm A}-
q}{2 k_{F}^{\rm A}+q} \right|
\\
\nonumber
&+&
D_{\rm B}
\log \left|\frac{2k_{F}^{\rm B}-
q}{2k_{F}^{\rm B}+q} \right|.
\label{fffreq}
\end{eqnarray}
where
\begin{equation}
D_{\rm A}=|M_{2k_{F}^{\rm A}}|^{2}
\frac{2 m_{\rm A}^{*}a}{\pi  M N_{c}k_{F}^{\rm A}}
\end{equation}
and
\begin{equation}
D_{\rm B}=|M_{2k_{F}^{\rm B}}|^{2} \frac{m_{\rm B}^{*}a}
{ \pi M N_{c} k_{F}^{\rm B}}
\end{equation}
All of the quantities needed to calculate the coefficients
$D_{\rm A}$ and $D_{\rm B}$have been obtained already.
We assume that the  bare phonon frequencies
are fit well by the form $(\Omega_{q\mu}^{0})^2=a_{0}+a_{1}q+a_{2}q^2$.
We then
use $a_{0} ,a_{1}$ and $a_{2}$ as fitting parameters to fit
our expression for $\Omega_q$ to the calculated FPA frequencies.
Doing this thereby enables us to extract the important bare frequency
dispersion $\Omega_q^0$ which is shown in Fig.~\ref{fig:pho5,0}.
Extracting these bare frequencies $\Omega_q^0$ allows us
to calibrate the effective Fr\"olich Hamiltonian Eq.~(\ref{frolich})
which will be used to study instabilities of the electron-phonon
system.  With our previously calculated quantities, we obtain
$D_{\rm A}=(219 $ cm$^{-1})^2$ and $D_{\rm B}=(146 $ cm$^{-1})^2$.
Using these values we thereby extract $\Omega_{q=2k_F^{\rm A}}^{0}=433$
cm$^{-1}$.

\subsection{(6,0) nanotube}
The band structure of the (6,0) CNT was considered
extensively by Blase \emph{et al}.~in Ref.~\onlinecite{Blase94}.  This tube
has a slightly larger diameter of 4.7 \AA.
The zone-folding band structure of this CNT is shown in the
left of Fig.~\ref{fig:6,0comb}.  As is typical of metallic zig-zag
tubes, there are two bands crossing at $k=0$ at the Fermi
energy.  The band structure directly computed with 24 atoms
in the unit cell is shown in the right of Fig.~\ref{fig:6,0comb}.
As discussed before \cite{Blase94}, these band structures
differ qualitatively which is a result of the hybridization
of the $sp_2$ and $p_z$ bands.  Here the inner band (with smaller
$k_F^{\rm B}$) is nondegenerate and originates from the $p_z$ bonds in
graphene while the outer band (with larger $k_F^{\rm A}$) is degenerate
and originates from the $sp_2$ bonds in graphene.

The coupling matrix elements for the (6,0) CNT were computed and
the coupling for the most dominant modes are shown in Fig.~\ref{fig:modes}
and Table \ref{table:modes6,0}.
The dominant inner band couplings were for intraband processes and are,
in descending order, to the out-of-plane optical and an in-plane
optical.
The dominant outer band couplings processes were found to
be the out-of-plane optical mode, an in-plane optical mode, the
radial breathing mode, and the in-plane acoustic stretching mode.

\begin{figure}
\includegraphics[width=3in]{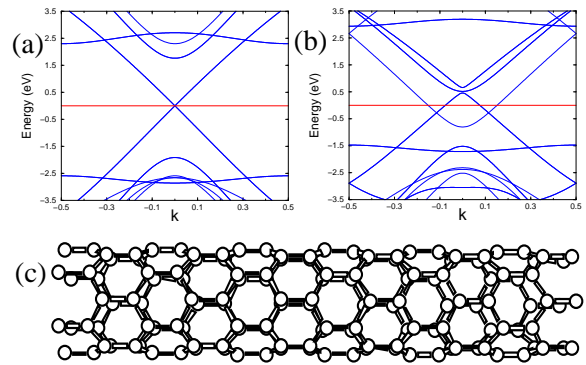}
\caption{The band structure of the (6,0) CNT obtained through
zone-folding (a) and calculated directly (b) along with the
atomic structure (c).}
\label{fig:6,0comb}
\end{figure}

\begin{table}
\begin{tabular}{c|c|c|c|}
(6,0) & mode & $\omega^{graph}_{q}$  (cm$^{-1}$)& $M_{kk'}$(eV/\AA)\\ \hline
$2k_{F}^{\rm A}$&4&857&7.27\\
&5&1585&6.80\\

\hline
$2k_{F}^{\rm B}$&4&847& 6.84\\
&6&  1591 & 6.12\\
&3&68&3.73\\
&2&493&2.31\\

\end{tabular}

\caption{Calculated values for the dominant coupling processes
for the (6,0) CNT.  The numbering scheme here
corresponds to that given in Fig.~\ref{fig:modes}.
$2k_{F}^{\rm A}$ and $2k_{F}^{\rm B}$ correspond to inner and
outer band processes respectively.  Phonon frequencies are given for
graphene.}
\label{table:modes6,0}
\end{table}

\begin{figure}
\includegraphics[width=3in]{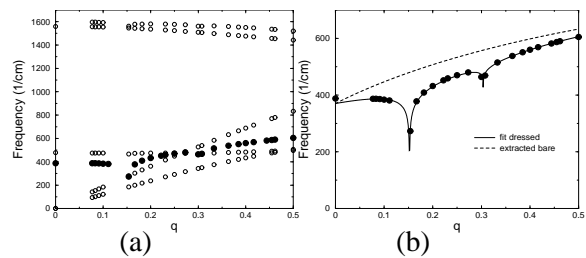}
\caption{(a): Phonon dispersion for the (6,0) CNT along the
$\Gamma M$ line of graphene.
(b):  The mode showing the most softening fit to the RPA
expression.}
\label{fig:pho6,0}
\end{figure}

Using the same procedure as was used for the (5,0) CNT in
the previous section for extracting the bare phonon frequency
at $2k_F^{\rm A}$.  From the previously computed values for
the electron-phonon coupling matrix elements and the band structure,
we find
$D_{\rm A}=(166 $ cm$^{-1})^2$ and $D_{\rm B}=(107 $ cm$^{-1})^2$.
After fitting, we extract the value  $\Omega_{q=2k_F^{\rm A}}^{0}=480$
cm$^{-1}$.

\subsection{(5,5) nanotube}

Finally, we study the more conventional armchair (5,5) CNT which
has a diameter of around 6.8 \AA.  As shown in Fig.~\ref{fig:5,5comb},
the zone-folding and directly computed band structure for this
larger diameter tube agree quite will.  Both of these band structures
show two bands which originate from $p_z$ orbitals which cross
at the Fermi energy at around $k=\frac{2}{3}\frac{a}{\pi}$.

\begin{figure}
\includegraphics[width=3in]{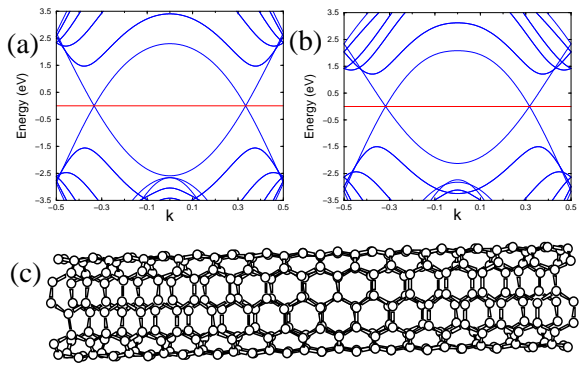}
\caption{The band structure of the (5,5) CNT obtained through
zone-folding and calculated directly are shown in the upper right
and left.  Bottom: the structure of the (5,5) CNT.}
\label{fig:5,5comb}
\end{figure}

The largest couplings for the CNT were found to again be from
the intraband processes and are shown in Fig.~\ref{fig:modes}
and Table \ref{table:modes5,5}.  The only significant intraband
coupling is for an in-plane mode
shown denoted by 7 in Fig.~\ref{fig:modes}.  The wave vector
for this mode is at the $K$ point in the first Brillouin zone of graphene.
For the interband processes, there is coupling
to the the radial breathing mode, but this is significantly
smaller.

\begin{table}
\begin{tabular}{c|c|c|c|}
(5,5) & mode & $\omega^{graph}_{q}$  (cm$^{-1}$)& $M_{kk'}$(eV/\AA)\\ \hline
&7&1479&11.60\\
&3&542&4.64\\
\end{tabular}

\caption{Calculated values for the dominant coupling processes
for the (5,5) CNT.  The numbering scheme here
corresponds to that given in Fig.~\ref{fig:modes}.
 Phonon frequencies are given for
graphene.}

\label{table:modes5,5}
\end{table}

\begin{figure}
\includegraphics[width=3in]{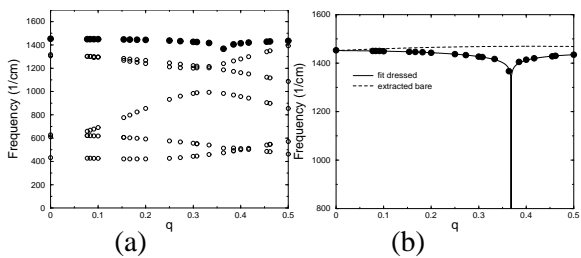}
\caption{(a): Phonon dispersion for the (5,5) CNT along the
$MK$ line of graphene.
(b):  The mode showing the most softening fit to the RPA
expression.}
\label{fig:pho5,5}
\end{figure}

For the (5,5) CNT, applying our method of extracting the
bare phonon frequencies, we obtain
$D_{\rm B}=(228 $ cm$^{-1})^2$.  Note that for this system,
only $\pi$ bands are relevant at the Fermi surface.  We
extract
$\Omega_{q=2k_F^{\rm B}}^{0}=1469$ cm$^{-1}$.

It is worth pointing out that there
has been some controversy about the relevant
phonon mode which couples the electrons at the Fermi surface for the
(5,5) CNT.\cite{Huang96,Caron00}  Our results confirm the 
study of Ref.~\onlinecite{Caron00}. The $2k_{F}$
processes couple to the phonons at the K point of graphene and the
relevant graphene mode has polarization vectors
$\hat{\epsilon}_{q}(1)=\frac{1}{\sqrt{2}}(i,1,0)$ and
$\hat{\epsilon}_{q}(2)=\frac{1}{\sqrt{2}}(1,i,0)$.  This out-of-phase
circular motion is qualitatively different from the linear
oscillations thought to couple previously.

\section{INSTABILITIES OF THE ELECTRON-PHONON SYSTEM}
\label{sec:instabilities}

\subsection{Charge-Density Wave Order}
\label{sec:CDW}

The RPA analysis presented in Sec.~\ref{sec:frequencies} can be
used to investigate the CDW (Peierls) transition temperature. This
instability corresponds to softening of the phonon frequency to zero, so
we can obtain it from the condition $\Omega_{Q_\tau}=0$ in
Eq.~(\ref{poles}) where $Q_\tau=2k_{{\rm F}\tau}$ is one of the
nesting wave vectors of the Fermi surface. The electron polarization
evaluated at temperature $T$  is given by
$
\chi_{0\tau}(2k_{{\rm F}\tau}, \omega=0,T)
= \frac{1}{2}\nu_\tau (0) \log
(T/4\varepsilon_{{\rm F}\tau})
$,
where $\nu_\tau (0) =2 m^*_\tau/Lk_{{\rm F}\tau}$ is the
contribution to the total density of states from band $\tau$.
We introduce the CDW coupling constant
\begin{eqnarray}
\label{lambdaCDW}
\lambda_{{\rm CDW},\tau\mu} =
\frac{|g_{Q_\tau\mu}|^2 \nu_\tau (0)}
{\Omega_{Q_\tau\mu}^0}
\end{eqnarray}
where $\tau$ specifies which of the $2k_F$ nesting
wave vectors
we are considering and $\mu$ labels the phonon mode.
Note, that distinguishing between various phonon modes
is important, since it tells us about the
nature of the distortion of atoms below
the Peierls transition (i.e. the in the plane vs
out of the plane).  One finds for the CDW transition
temperature
\begin{eqnarray}
\label{CDWTC}
T_{{\rm CDW},\tau\mu} = 4 \varepsilon_{{\rm F}\tau}
e^{-1/\lambda_{{\rm CDW}\tau\mu}}.
\end{eqnarray}
Corrections to this equation due to an additional band with
different Fermi wave vector (e.g. the term with the logarithmic
divergence at 2$k_{F}^{B}$ in 
Eq.~\ref{Eq:fit}) is small and will be neglected.
Degenerate bands
(e.g.~the $\rm A$ band for the (5,0) CNT), are accounted
for by an additional factor of 2 in the density of
states is Eq.~(\ref{lambdaCDW}).  In Table \ref{table:CDW}
we summarize our results for the CDW instability for
the CNTs studied.

\begin{table}
\begin{tabular}{l|c|c|c|}
&(5,0)&(6,0)&(5,5)\\ \hline
mode &4&4&7 \\
$\Omega_{2k_{F}}^{0}$(cm$^{-1}$) &433 &480&1469 \\
$\lambda_{\rm CDW}$&0.26&0.12&0.024 \\
$T_{\rm CDW}$ (K)&160&5& $7\times 10^{-14}$\\
\end{tabular}
\caption{The dominant mode for the CDW instability,
the extracted bare phonon frequency, the CDW
coupling parameter, and the CDW transition temperature
for the various CNTs studied.}
\label{table:CDW}
\end{table}

\subsection{Superconductivity}
\label{sec:SC}

To analyze the superconducting instability of the
CNTs we use the Migdal-Eliashberg theory.
The isotropic Eliashberg equations
for the one-dimensional case, neglecting the Coulomb interaction,
can be written as (see Appendix \ref{Appendix:Migdal-Eliashberg}
for details)
\begin{equation}
\label{eli1}
Z_{n}=1+f_{n}s_{n}\sum_{n'}\lambda(n-n')s_{n'}
\end{equation}
\begin{equation}
\label{eli2}
Z_{n}\Delta_{n}=\sum_{n'} \lambda(n-n') f_{n'}\Delta_{n'}
\end{equation}
where $f_{n}=1/|2n+1|$, $s_{n}={\rm sgn}(2n+1 )$, $\Delta_{n}=\phi_{n}/Z_{n}$, and
the frequency dependent coupling constant $\lambda(n)$ is given by
\begin{eqnarray}
\nonumber
\lambda(n-n')=-\frac{1}{\nu_{\sigma}(0)}&\sum_{k\tau k'\tau' \mu}&
\delta(\varepsilon_{k\tau})\delta(\varepsilon_{k'\tau'})|g_{kk'\mu}|^{2}
\\
&\times&D_{\mu}(k-k',n-n')
\label{eq:LambdaFreq}
\end{eqnarray}
where $\nu_{\sigma}(0)$ is the density of states per spin at
the Fermi energy.
When analyzing superconductivity in two and three
dimensional systems using  the Eliashberg equations
it is sufficient to take the
bare phonon  propagators
$D_0(k-k',n-n')$
in Eq.~(\ref{eq:LambdaFreq}). This is justified since in the
absence of Fermi surface nesting there is typically little difference
between the bare and the dressed phonon frequencies and propagators.
In one-dimensional systems, however, there is a strong temperature dependent
renormalization of the phonon spectrum which needs
to be taken into account.
The simplest way to do so is to use the
FPA form of the phonon propagator
(see eqs (\ref{eq:Dyson}) - (\ref{Eq:polesFPA}))
\begin{eqnarray}
D_{\mu}^{\rm FPA}(q,i\nu_m)=\frac{2\Omega^0_{q\mu}}{(i\nu_m)^2-
(\Omega_{q\mu})^2}
\end{eqnarray}
Here $\Omega_{q\mu}$ is the dressed phonon
frequency in the FPA given in Eq.~(\ref{Eq:polesFPA}).
Taking a soft dressed phonon propagator immediately leads to
the enhancement of the electron pairing via the increase of
$\lambda(n)$. Enhancement of superconductivity by the giant Kohn
anomaly in one-dimensional systems has been discussed previously by
Heeger in Ref.~\onlinecite{Heeger79}. The main
subtlety of the Eliashberg equations in this case is that the
phonon frequency $\Omega_{q\nu}$ now has temperature dependence
which needs to be found using the finite temperature form of the polarization
operator $\Pi(q,0)$ in Eq.~(\ref{Eq:polesFPA}).

When we analyze the (5,0) nanotube following this strategy, we find,
however, that the CDW instability always appears before the
superconducting one. This is in agreement with the general argument
proposed in Ref.~\onlinecite{Ginzburg82} that in strictly one-dimensional
electron-phonon systems Peierls instability alway dominates, since it
involves all electrons in the band, compared to the superconducting
instability, which involves only electrons in the vicinity of the
Fermi surface.

To introduce a quantitative measure of the strength of
superconducting pairing we use the {\it bare}
phonon propagator in Eq.~(\ref{eq:LambdaFreq}).
This approximation will be more carefully considered
in Sec.~\ref{sec:SCren}, along with inclusion of the
Coulomb interaction.
A useful approximate solution of the Eliashberg
equations (\ref{eli1}) - (\ref{eq:LambdaFreq})
is given by the McMillan formula (again in the absence
of Coulomb interaction) \cite{McMillan68,Allen82}
\begin{equation}
T_{\rm SC}=\frac{\langle \Omega \rangle}{1.20}
\exp \left[- \frac{ 1.04(1+\lambda_{\rm SC})}{\lambda_{\rm SC}} \right]
\end{equation}
Here $\lambda_{\rm SC}$ is the zero frequency component
of Eq.~(\ref{eq:LambdaFreq}) where, again, the bare phonon
frequencies are used
\begin{equation}
\label{eq:LambdaZero}
\lambda_{\rm SC}=-\frac{1}{\nu_{\sigma}(0)}\sum_{kk'}
\delta(\varepsilon_{k})\delta(\varepsilon_{k'})|g_{kk'}|^{2} D_0(k-k',0).
\end{equation}
In accordance with Ref.~\onlinecite{Benedict95}, we take
$\langle \Omega \rangle= 1400$ K.  The superconducting
coupling constants and transition temperatures calculated
in this manner are summarized in Table \ref{table:SC}.
We emphasize, however, that these numbers should be taken with
some scepticism, since within the same approximation
the CDW instability is usually the dominant one and appears
at much higher temperatures (compare to Table \ref{table:CDW}).

\begin{table}
\begin{tabular}{l|c|c|c|}
&(5,0)&(6,0)&(5,5)\\ \hline
$\lambda_{\rm SC}$&0.57&0.12&0.031 \\
$T_{\rm SC}$ (K)&64&0.071& $1.11\times 10^{-12}$\\
\end{tabular}
\caption{The SC
coupling parameter, and the SC transition temperature
for the various CNTs studied. The CDW instability and the residual Coulomb
interaction between electrons are neglected in the calculation
of these quantities.}
\label{table:SC}
\end{table}

Finally, it is known that $q \approx 0$ scattering processes
due to acoustic phonons
can be important in one-dimensional 
electron-phonon 
systems.\cite{Wentzel51, Bardeen51, Engelsberg64}  However, in the
approximations leading to Eq.~\ref{eq:LambdaZero} these contributions
were neglected.  In Appendix \ref{Appendix:Acoustic}
we show that while these
processes can be important for some systems, their inclusion
leads to only a small correction to $\lambda_{\rm SC}$ for the
CNTs we study.  This is due to the fact that the dominant contributions
to the superconducting coupling constant are from optical phonons.

\section{Role of the Coulomb interaction}
\label{sec:coulomb}

In the discussion above we concentrated
on the electron-phonon interaction
with electron-electron Coulomb interaction
included only at the mean-field level
via the band structure.
It is useful to consider how the residual
Coulomb interaction can modify the
analysis of the Peierls and superconducting instabilities
discussed above. We take
\begin{eqnarray} 
\label{Eq:CH}
{\cal H} &=& {\cal H}_{\rm e-ph}+{\cal H}_{\rm e-e}
\\
{\cal H}_{\rm e-e} &=\frac{1}{2}& \sum_{kk'q\tau \tau' \sigma\sigma'}
V_{q\tau \tau'} c^\dagger_{k+q \tau \sigma} c^\dagger_{k'-q \tau' \sigma'}
 c_{k' \tau' \sigma'} c_{k \tau \sigma}
\nonumber
\end{eqnarray}
where ${\cal H}_{\rm e-ph}$ is still given by Eq.~(\ref{frolich}) and
we will always assume $k$ and $k'$ around the Fermi surface.  Note that
we have neglected interband scattering
 which is typically small.
In the following, we will consider how introducing this Coulomb
interaction modifies the results. 

\subsection{Coulomb interaction potential}
For the Coulomb interaction between conduction electrons, we
take the form used by Egger \emph{et al}.~in
Ref.~\onlinecite{Egger98}
\begin{equation}
\label{Eq:EggerCoul}
V({\bf r-r'})=\frac{e^2/\kappa}{\sqrt{(x-x')^{2}+
\left(2R \sin\left(\frac{y-y'}{2R}\right)\right)^{2}+a_{z}^{2}}}.
\end{equation}
Here, the $y$-direction is chosen to be along the perimeter of the
CNT and $x$ measures the distance along the CNT axis.
A measure of the spatial extent of the
$p_{z}$ electrons perpendicular to the CNT is given by
$a_{z} \approx 1.6$ \AA~and $R$ is the CNT radius.  Note that this interaction potential is periodic in
the $y$-direction.
For the dielectric constant due to the bound electrons, we will take
the value $\kappa \approx 2$ predicted by the model
of Ref. \onlinecite{Benedict95_2}.

We can now use Eq.~(\ref{Eq:EggerCoul}) to obtain the
Coulomb interaction entering Eq.~(\ref{Eq:CH})
\begin{eqnarray}
\label{Eq:Coulombintegral}
V_{q\tau\tau'}=
&\int& d^{2}r d^{2} r' V({\bf r}-{\bf r}')
\\
\nonumber
&\times &
\psi^{*}_{k+q\tau}({\bf r})
\psi_{k\tau}({\bf r}) \psi^{*}_{k'-q\tau'}({\bf r}')\psi_{k'\tau'}({\bf r}').
\end{eqnarray}
The region of integration above is over areas of length $L$
along the $x$-direction where $L$ is the length of the system 
and of width $2 \pi R$ along the $y$-direction.
For backward
scattering  processes ($q \approx 2k_{F}$)
between the inner bands  of the (5,0) and (6,0) CNTs we find that
$V_{q\tau\tau'}$ is independent of $\tau$ and $\tau'$, and
(see Appendix \ref{Appendix:CoulDerivation} for derivation)
\begin{eqnarray}
\label{Eq:BSCoulomb}
V_{q}&\approx& \gamma
\frac{1}{L^{2}}\int dx dx' e^{-iq(x-x')}
\nonumber
\\
&\times& \int_{0}^{2 \pi R}\frac{dy}{2\pi R}\int_{0}^{2 \pi R}\frac{dy'}{2\pi R}
V({\bf r}-{\bf r}')
\end{eqnarray}
where $\gamma=$ 0.59 and 0.0016 for the (5,0) and
(6,0) CNTs respectively.  This is significantly reduced from the value
of $\gamma \approx 1$ that one obtains for larger radius 
CNTs \cite{Egger98} which is due to the fact that wave functions
at the Fermi points have  different symmetries for the (5,0) CNT
and (6,0) CNTs.
More specifically, it can be found that 
$\mu_{\rm CDW}$ is very 
small for the (6,0) CNT due to the fact that for metallic 
zig-zag nanotubes, the wave functions at $-k$ and $k$ close to the 
Fermi energy are nearly orthogonal within the unit cell of the CNT since they
correspond to symmetric and antisymmetric combinations of
atomic orbitals in the graphene sheet.

\subsection{Modification of CDW instability due to Coulomb interaction}
\label{sec:CDW_Coulomb}

\begin{figure}
\includegraphics[width=3in]{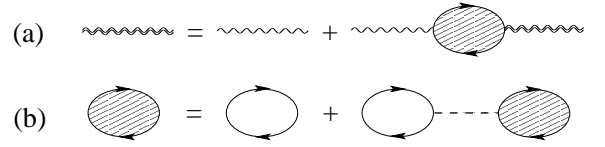}
\caption{Dyson's equation for the phonon propagator (a) where the
Coulomb interactions are taken into account within the RPA (b).}
\label{fig:phoelecRPA}
\end{figure}

The simplest approximation (beyond mean-field)
which includes the Coulomb repulsion is the RPA
shown in Fig.~\ref{fig:phoelecRPA}
(see e.g. Refs.~\onlinecite{Schrieffer64, Levin74}),
Eq.~(\ref{poles}) now becomes for a one-band system
\begin{equation}
\label{polesWITHcoulomb}
\left( \Omega_{q\mu}\right)^{2}=\left( \Omega_{q\mu}^{0}\right)^{2}
+2\Omega^0_{q\mu}\frac{\Pi_{\mu}(q,\Omega_{q\mu})}
{1-V(q)\chi_{0}(q,\Omega_{q\mu})}
\end{equation}
where $\Pi_{\mu}(q,\Omega_{q\mu})=|g_{q\mu}|^{2}\chi_{0}(q,\Omega_{q\mu})$.
We immediately see that including the  Coulomb interaction
can suppress the CDW instability.
The second term in Eq.~(\ref{polesWITHcoulomb}) no longer
diverges when $q=2k_F$ and the softening of the
$2k_F$ phonons occurs only for
$\mu_{{\rm CDW}} <
\lambda_{{\rm CDW},\mu}$, where
\begin{equation}
\mu_{{\rm CDW}} \equiv \frac{1}{2} \nu(0) V_{q=2k_{F}}.
\end{equation}
From equation
(\ref{polesWITHcoulomb}) we also find how
the Coulomb interaction modifies the Peierls
transition temperature
\begin{equation}
\label{Eq:TCDWwithCoulomb}
T_{{\rm CDW},\mu} = 4 \varepsilon_{{\rm F}} \exp\left(
-\frac{1}{[\lambda_{{\rm CDW},\mu}-\mu_{{\rm CDW}}]} \right).
\end{equation}

We will now estimate the magnitude
of $\mu_{\rm CDW}$ from this residual Coulomb
interaction for the (5,0) which was seen above to be the most unstable
toward the formation of a CDW from distortion of the out-of-plane
optical mode shown in Fig.~\ref{fig:modes}.
Carrying through the straightforward generalization of the RPA
analysis for the multiple-band system, and carrying out the integrals in
Eq.~(\ref{Eq:BSCoulomb}) for the Coulomb backward scattering
interaction, we obtain $\mu_{\rm CDW}=0.24$.
Note that this is quite close to $\lambda_{\rm CDW}=0.26$ for
this particular instability.
This indicates that it is possible that the Coulomb interaction
can significantly suppress the CDW transition temperature or even
remove the CDW instability altogether.  Indeed, taking these values we
find that $T_{\rm CDW}$ is suppressed to less than $10^{-18}$ K.

For the (6,0) CNT, we calculate the smaller value
$\mu_{\rm CDW}=0.0019$.  This will not change the
value of $T_{\rm CDW}=5.0$ K that we calculated previously
for the (6,0) CNT.

\subsection{Phonon vertex renormalization through screening}
\label{sec:vertex_Coulomb}

It can be seen that the Coulomb interaction further can screen
 the electron-phonon
vertex.  By including screening through the RPA,
 we find that the screened vertex
is given by \cite{Schrieffer64}
\begin{equation}
\bar{g}_{q\mu}=\frac{g_{q\mu}}{1-V_{q}\chi_{0}(q)}
\end{equation}
for a one-band system where
$\chi_{0}(q) \equiv \chi_{0}(q,\Omega_{q\mu}=0)$.
Thus we see that the inclusion of screening reduces the electron-phonon vertex.
We note that in the treatment in Sec.~\ref{sec:CDW_Coulomb} of the CDW
instability it would
be inappropriate to use the screened vertices since this would lead to
double-counting.

Full charge self-consistent calculations will determine the dressed
electron-phonon vertex (see Appendices
\ref{Appendix:Mkk'} and \ref{Appendix:SCC}).
This is desirable in 3d, where the renormalization is presumably small.
 However in 1d, one would calculate greatly suppressed values for the
couplings, dominated by the screening due to the logarithmic
divergence of the susceptibility at $2k_{F}$.
Because of the subtle interplay
between these divergences, it is  desirable to
calculate the bare vertex and then manually
put in the Coulomb interaction as
we do.

Since with the method we use, the charge distribution is not
calculated self-consistently, we calculate the
bare electron-phonon vertex $g_{q\mu}$.  We point out, however, that there
is an approximation here.  The true bare electron-phonon vertex should
be calculated in the absence of the conduction electron entirely which
is separately accounted for in the residual Coulomb term.
In our method, however,
the conduction electron is taken to
adiabatically follow the ion through the distortion.
Because of this, we expect our results to slightly underestimate the bare
electron-phonon coupling vertices.

\subsection{Modification of superconducting instability
due to Coulomb interactions}
\label{sec:SCren}

To include the Coulomb interaction in the Eliashberg equations,
it is necessary to dress both electron-phonon vertices shown in
Fig.~\ref{fig:migdal} according to
Sec.~\ref{sec:vertex_Coulomb} as well as the
phonon propagator according to Sec.~\ref{sec:CDW_Coulomb}.
This leads to the modified phonon-mediated interaction between electrons
of
\begin{eqnarray}
|\bar{g}_{q\mu}|^{2}D_{\mu}(q,\Omega)
&=&\frac{|g_{q\mu}|^{2}}{(1-V_{q}\chi_{0}(q))^{2}}
\\
&\times&\frac{2\Omega_{q\mu}^{0}}
{\Omega^{2}-(\Omega_{q\mu}^{0})^{2}-2\Omega_{q\mu}^{0}|g_{q\mu}|^{2}\frac{\chi_{0}(q)}{1-V_{q}\chi_{0}(q)}}.
\nonumber
\end{eqnarray}
Using this leads to a modified result for the
superconducting coupling constant $\lambda_{\rm SC}$.
For a specific process of wave vector $q$, coupling
 points on the Fermi surface,
we find that the renormalized contribution to 
the superconducting coupling constant is
given by
\begin{equation}
\label{lambdaren}
\lambda_{q\mu}=\left(\frac{1}{(1-V_{q}\chi_{0}(q))^{2}}\right)\left(\frac{1}
{1+\frac{2|g_{q\mu}|^{2}}{\Omega_{q\mu}^{0}}\frac{\chi_{0}(q)}{1-V_{q}\chi_{0}(q)}}\right)\lambda_{q\mu}^{0}
\end{equation}
where $\lambda_{q\mu}^{0}$ is the unrenormalized contribution.
All such contributions
must be summed over to determine the total $\lambda_{\rm SC}$.
The first and second factors tend to decrease and increase
the electron-phonon coupling
respectively.  Physically, the first factor is due the screening of the electron-ion interaction
due to conduction electrons.  The second factor is due to the softening of particular modes
due to the Kohn Anomaly which will in turn enhance the overall electron-phonon coupling.
Since these renormalization factors
depend on temperature through the susceptibility $\chi_{0}$,
$T_{\rm SC}$ must  be determined self-consistently.

In addition to the renormalization of the Coulomb vertex, there is
also the direct Coulomb repulsion between electrons that is
taken into account through the Coulomb pseudopotential $\mu^{*}_{\rm SC}$
which is included in McMillan's expression \cite{McMillan68,Allen82}
\begin{eqnarray}
\label{Eq:McMillan}
T_{\rm SC}=\frac{\langle \Omega \rangle}{1.2}
\exp\left(
-\frac{1.04(1+\lambda_{\rm SC})}{[\lambda_{{\rm SC}}-\mu^*_{{\rm SC}}
(1+0.62\lambda_{\rm SC})
]} \right)
\end{eqnarray}
where
\begin{eqnarray}
\label{Eq:mustar}
\mu^*_{{\rm SC}}=\frac{\mu_{{\rm SC}}}
{1+\mu_{{\rm SC}} \ln\left(\frac{E_{F}}{\omega_D}\right)}
\end{eqnarray}
and $\mu_{\rm SC}$ is the screened Coulomb interaction averaged over the
Fermi surface.

We will now estimate $\mu^{*}_{\rm SC}$.  Taking into account
screening within the RPA one finds
\begin{equation}
V^{s}_{q}=\frac{V_{q}}{1-V_{q}\chi_{0}(q)}
\end{equation}
for the screened Coulomb interaction.
In 1d for $q \approx 2k_{F}$, $V^{s}_{q}\approx 0$.  This is due to the divergence
of $\chi_{0}(q)$ at $q=2k_{F}$.  Also, one finds that for $q \approx 0$,
$V^{s}(q) \approx 1/\nu(0)$.  Using this RPA screened Coulomb interaction
we find for our three band system of the (5,0) CNT
\begin{equation}
\mu_{\rm SC} \equiv \frac{1}{\nu_{\sigma}(0)}\sum_{k\tau k'\tau'}\delta(\varepsilon_{k\tau})
\delta(\varepsilon_{k'\tau'})V_{\tau\tau'}^{s}(k-k')=0.25.
\end{equation}
Then, using Eq.~(\ref{Eq:mustar}), we obtain $\mu_{\rm SC}^{*}=0.19$ for
the Coulomb pseudopotential with the calculated values of the Fermi
energy and Debye frequency.

We now see how taking into account the Coulomb interaction
in this manner modifies the superconducting transition temperature
for the (5,0) CNT.
The most significant renormalization of the total superconducting $\lambda_{\rm SC}$
given by Eq.~(\ref{lambdaren}) will be for
the $2k_{F}$ process that couples to the out-of-plane optical
mode which was previously seen to have the overall strongest
coupling.  That is, at temperatures where
the renormalized
$\lambda_{\rm SC}$ will start to differ from the bare
$\lambda_{\rm SC}^{0}$,
all of the renormalization will come from this mode.
Using Eqns.~(\ref{lambdaren})
and \ref{Eq:McMillan}, we find that a self-consistent solution for the
superconducting transition temperature occurs
at $T_{\rm SC}\approx 1.1$ K which
is larger than the previously calculated CDW transition temperature.  This
therefore shows that the Coulomb interactions can
favor superconductivity over the CDW instability.

For the (6,0) CNT, we see that the value of $T_{\rm SC}$ 
without the inclusion Coulomb interaction is smaller
than $T_{\rm CDW}=5$K that we computed in the previous section
with the inclusion of the Coulomb interaction.  We therefore conclude that
the CDW will be dominant for the (6,0) CNT.

\subsection{Summary of Coulomb effects}
In the above, we have shown that the introduction of the residual Coulomb
interaction will lower both the SC and CDW transition temperatures.  We
also illustrated the possibility that the CDW instability can be suppressed
so much by Coulomb interactions that SC will be dominant
at low temperatures.  However, we stress the difficulty
of obtaining such quantitative results.  In principle, to obtain
an accurate Coulomb interaction in our basis of Bloch states, one
needs to use the interaction
\begin{equation}
V_{kk'k''k'''}=\frac{1}{\kappa}\int d^{3} r d^{3} r'\psi_{k}^{*}(\textbf{r})\psi_{k'}^{*}(\textbf{r}')
\frac{e^{2}}{|\textbf{r}-\textbf{r}'|}
\psi_{k''}(\textbf{r}')\psi_{k'''}(\textbf{r})
\end{equation}
where the $\psi_{k}$'s are Bloch state wave functions of the CNT which is
difficult to obtain.
The Coulomb interaction $V_{q}$ we used is only
a rough approximation to this more realistic interaction.
On the other hand, the SC and CDW transition temperatures have exponential
dependence on the Coulomb interaction parameters.  One also
has to be very careful not to double-count the electron-electron interaction
terms taken into account in the single-particle energies $\varepsilon_{k}$
through the Hartree term.  As shown in Appendix \ref{Appendix:SCC},
using a method in which the charge density is calculated
self-consistently will give more accurate
values for the phonon frequencies calculated through the FPA.  However,
there are serious difficulties with calculating 
the bare electron-phonon vertex with such a 
method as discussed in Sec.~\ref{sec:vertex_Coulomb}.

\section{DISCUSSION}
\label{sec:discussion}

\subsection{Comparison to other carbon based materials}
As a consistency check, we now compare our results
for the attractive potential due to electron-phonon coupling
to established
results for other carbon based solids, namely the intercalated
graphene KC$_8$ and the carbon fullerene K$_3$C$_{60}$.
Calculations of the density of states at the Fermi energy
yield $\nu(0)=$ 0.24 (Ref.~[\onlinecite{Inoshita77}])
and 0.29 (Ref.~[\onlinecite{Antropov92}])
states/ eV / C atom for KC$_8$ and K$_3$C$_{60}$ respectively.
Estimates of $\lambda_{\rm SC}$ for these are 0.21
(Ref.~[\onlinecite{Inoshita77}]) and
0.7 (Ref.~[\onlinecite{Benedict95}]).
In the BCS theory, $\lambda_{\rm SC}$ is expressed in terms of the
product of the electronic density of states at the Fermi level
and the attractive pairing potential strength $\lambda_{\rm SC}=\nu(0)V$.
\cite{Schrieffer64}  Now that we have the magnitude $\lambda_{\rm SC}$
and $\nu(0)$, we can extract the magnitude of the pairing potential
for the intercalated graphene, the fullerene, and the CNTs we study.
The results are summarized in Table \ref{table:attractive}.

\begin{table}
\begin{tabular}{c|c|c|c|c|c|}
&KC$_8$&K$_3$C$_{60}$&(5,0) CNT & (6,0) CNT & (5,5) CNT\\ \hline
$\nu(0)$ (eV$^{-1}$)&0.24$^a$&0.29$^b$&0.16&0.068&0.034\\
$\lambda_{\rm SC}$&0.21$^a$& 0.7$^c$ & 0.57 & 0.12 & 0.031 \\
$V$ (eV) & 0.875 & 2.4 &3.6 & 1.8 & 0.92 \\
\end{tabular}
\caption{Density of states at the Fermi energy, the superconducting
coupling strength, and the attractive potential strength for various
carbon materials. Superscripts
$a,b,$ and $c$ denote Refs.~[\onlinecite{Inoshita77}],
[\onlinecite{Antropov92}], and [\onlinecite{Benedict95}] respectively.
}
\label{table:attractive}
\end{table}

The following analysis
will be very similar to that of Benedict \emph{et al}.~in
Ref.~\onlinecite{Benedict95}
The central idea in their analysis is as follows.  Since curvature
increases the amount of hybridization between $\sigma$ and
$\pi$ states at the Fermi energy, the strict selection rules
for phonon scattering between pure $\pi$ states in graphene
will be lifted.  The amount of $\sigma-\pi$ hybridization has
roughly a $1/R$ dependence on the radius of curvature, so the matrix
elements and therefore the attractive
potential due to curvature will go as $1/R^2$.

\begin{figure}
\includegraphics[width=3in]{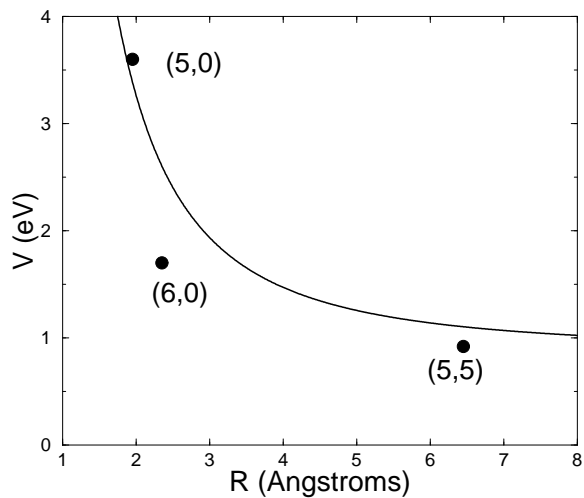}
\caption{$V_{\rm tube}(R)$ from Eq.~(\ref{V_fit}) calibrated with
parameters from intercalated graphene and fullerenes (solid line)
compared to the attractive potentials calculated for the representative
CNTs (filled circles).}
\label{fig:attractive}
\end{figure}

Neglecting presence of pentagons in fullerenes, we write
the attractive potential for the fullerene $V_{\rm{ball}}$ as
the sum of contributions from that of the graphene sheet
$V_{\rm{flat}}$ and that from curvature effects $V_{\rm{curve}}$

\begin{equation}
V_{\rm{ball}}=V_{\rm{flat}}+V_{\rm{curve}}.
\end{equation}
This relation enables us to obtain the value for
$V_{\rm{curve}}=1.5$ eV.
Now we can write the expected attractive interaction
for the CNT

\begin{equation}
V_{\rm{tube}}(R)=V_{\rm{flat}}+V_{\rm{curve}}
\left(\frac{R_0/2}{R}\right)^{2}
\label{V_fit}
\end{equation}
where $R_{0} \approx 5$ \AA~is the radius of a fullerene and
the factor of two comes in because there is twice as
much $\sigma-\pi$ hybridization in a fullerene as there
is in a CNT of radius $R_{0}$. \cite{Benedict95}
In Fig.~\ref{fig:attractive},
we show that Eq.~(\ref{V_fit}), which was calibrated by using
only quantities from intercalated graphene and fullerenes,
is consistent with the attractive potentials
we obtain for the (5,0), (6,0), and (5,5) CNTs.

\subsection{Beyond Mean Field Theory}

One-dimensional electron-phonon systems have
several competing instabilities and the true ground 
state may be found only by analyzing 
their interplay.~\cite{Voit87,Voit88}  
Hence, one may be concerned that we use a mean-field approach to analyze
a 1d CNT. 
We point
out that when we calculate the superconducting $T_{\rm SC}$ we
include the interplay
of the CDW and SC orders.  That is, the effective superconducting
coupling $\lambda_{\rm SC}$ that we obtain in Eq.~\ref{lambdaren} includes
softening of the $2k_{F}$ phonon mode.  Such an approach is equivalent
to the two parameter RG analysis used in Ref.~\onlinecite{Grest76}.
The mean-field transition temperature obtained by our method is 
equivalent to the coupling constants becoming of the order of unity in the RG
analysis.   At $T_{\rm SC}$ electrons start to pair, but the system has
strong fluctuations in the phase of the SC order parameter.  The most
important kind of fluctuations are thermally activated phase slips,
discussed originally for superconducting wires in 
Refs.~\onlinecite{Langer67,McCumber70}.
Phase slips lead to only a gradual decrease of resistivity with temperatures
below $T_{\rm SC}$.

For an incommensurate CDW, long range order may not appear at finite 
temperature either.  To understand the physical meaning
of the mean-field transition, we can introduce a Landau-Ginzburg
formalism.~\cite{caron94} 
Here we concentrate on the (5,0) and (6,0) CNTs which have  
three partially filled bands with Fermi points $k_{F}^A$ and
$k_F^B$ where the exact relation $2k_{F}^A=k_F^B$ is satisfied.
We introduce a complex order parameter $\Psi_1(x)$ related to the amplitude of
the lattice distortion as $q(x)=e^{2ik^A_Fx}\, \Psi_1(x)+
e^{-2ik^A_Fx}\, \Psi^*_1(x)$. At low temperature the free energy
is given by
$ F_\sigma[\Psi_1]=\int dx ( a |\Psi_1|^2+ b |\Psi_1|^4
+c|\frac{d\Psi_1}{dx}|^2 ). $
Below the mean-field transition temperature $T_{\rm CDW}$ we have
$a<0$ and the system develops an amplitude for the order parameter
$\Psi_1$. The phase of $\Psi_1$, however, is still fluctuating,
leading to short range correlations for the CDW order $\langle
\Psi_1(x) \Psi^*_1(0) \rangle \propto e^{-x/\xi(T)}$. Even at
$T=0$ we can have at best a quasi long-range order for $\Psi_1$
due to the incommensurate value of $2 k^A_F$. Lattice distortions
at $2 k^B_F$ can be included by introducing another complex field
$\Psi_2(x)$ that contributes $e^{2ik^B_Fx}\, \Psi_2(x)+
e^{-2ik^B_Fx}\, \Psi^*_2(x)$ to the distortion amplitude. The
relation $2 k^A_F= k^B_F$ implies that the free energy allows
coupling between $\Psi_1$ and $\Psi_2$ of the form $F_{\pi
\sigma}[\Psi_1,\Psi_2]= \gamma \int dx (\Psi^2_1
\Psi^*_2+\Psi^{*2}_1 \Psi_2)$, so when the amplitude of $\Psi_1$
is established, it will immediately induce the amplitude for
$\Psi_2$ (although none of the fields have a long-range order).
Appearance of such amplitudes should lead to a pseudogap state of
the system below  $T_{\rm CDW}$.~\cite{caron94} The dominant
contribution to electrical conductivity in a clean system would
then come from the Goldstone mode of the phase of the $\Psi$'s,
i.e$.$ sliding of CDWs (Fr\"{o}lich mode). Any kind of disorder
(e.g$.$ impurities or crystal defects), however, gives strong
pinning of the CDW phase and suppresses collective mode
contributions to transport. Therefore, we expect insulating
behavior of the low temperature resistivity in most experimentally
relevant circumstances if CDW is the dominant low temperature
phase.

\subsection{Experimental Implications}

Proximity induced\cite{Kasumov99, Morpurgo99} as well as  
intrinsic \cite{Kociak01, Tang01} superconductivity
has been experimentally observed in carbon nanotubes.
On the other hand, the CDW state, despite 
being endemic to quasi 1d systems has never been reported
for carbon nanotubes.
As we discuss above, one needs to have very small carbon nanotubes
to have electron-phonon interaction strong enough to make
either the CDW or the SC instabilities appear at 
experimentally relevant temperatures.  In this work we
address quantitatively both of these
instabilities.  Our main conclusion is that when we include Coulomb
interaction between
electrons, the CDW instability does no appear even for the ultrasmall
nanotubes, whereas the superconducting $T_{\rm SC}$ may be
in the few Kelvin range.

In the work by Kociak \emph{et al$.$} in Ref.~\onlinecite{Kociak01},
electronic transport through ropes of single-walled CNTs suspended between
normal metal contacts was measured.  The ropes are composed of several hundred
CNTs in parallel with diameters of the order 1.4 nm.
It was found that below 0.5 K, the resistance abruptly drops, an effect
which is destroyed by the application of an external magnetic 
field of order 1 T.  
The largest radius CNT we study is the (5,5) CNT, 
which was seen to be in the regime where zone-folding
is applicable.  For this CNT, we calculated $\lambda_{\rm SC}=0.031$, a
value far too small to support superconductivity at this temperature
even without the inclusion of the Coulomb interaction.  This
small value of $\lambda_{\rm SC}$ is consistent with the experimental
measurements of the electron-phonon coupling in CNTs of similar
diameter by Hertel \emph{et al$.$} in Ref.~\onlinecite{Hertel00}.
It is possible that the interactions between CNTs in the rope
play a tantamount role for superconductivity in the experiment
of Ref.~\onlinecite{Kociak01} as suggested
by Gonzalez in Ref.~\onlinecite{Gonzalez02}.  Another
possibility is that a small number of 
nanotubes in the rope have a small diameter.  For nanotubes
with a diameter of 4 \AA~we find superconducting $T_{\rm SC}$
in the 1 K range which would be consistent with these experiments.  
A small number of superconducting nanotubes could provide a 
short-circuiting in transport measurements or even induce 
superconductivity in other CNTs via the proximity effect.

In the experimental work of Tang \emph{et al$.$} in Ref.~\onlinecite{Tang01},
electrical transport was measured 
through a zeolite matrix containing single-walled
CNTs.  In the zeolite matrix, the CNTs are well-separated from each
other creating an idealized one-dimensional system.  The diameters
of the CNTs were determined to be approximately 4 \AA~by measuring
the radial breathing phonon mode frequency by Raman spectroscopy. 
The superconducting transition temperature for this system was
found to be 15 K from transport measurements.  In addition,
the Meissner effect was observed through the temperature
dependence of the magnetic susceptibility suggesting that 
the large currents observed in transport measurements are not 
from the sliding charge-density wave collective mode, but are indeed from
superconducting correlations.

The ultrasmall (5,0) CNT we study is the 
likely candidate structure for the
CNTs confined in the zeolite matrix in these experiments.
We find for this system that the electron-phonon coupling is very
strong.  We find in the  mean-field theory, neglecting Coulomb interactions, 
that $T_{\rm CDW}=160$ K and $T_{\rm SC}=64$ K, indicating that
the charge-density wave instability is stronger in this approximation. 
However, putting in the Coulomb interaction
as in Eq.~(\ref{Eq:CH}), the charge-density wave transition was suppressed
to very low temperatures, making superconductivity dominant
with $T_{\rm SC}=1$ K. 
Discrepancy between our calculated $T_{\rm SC}$ and the
experimentally observed 15 K should not be a reason for concern.
The superconducting transition temperature 
in Eq.~\ref{Eq:McMillan} is exponentially sensitive to the
strength of the Coulomb interaction, and our estimates of the latter are not
very accurate.

\section{Summary and Conclusions}

In this work, we have used the Fr\"{o}lich Hamiltonian written in
Eq.~(\ref{frolich}) to study three types of small-radius CNTs.  For this
Hamiltonian, the band structure energies were computed by using an
empirical tight-binding method \cite{Mehl96} to first relax the
structure, and then to compute the eigenvalues of the secular
tight-binding equation.  The
electron-phonon interaction $g_{k\tau k' \tau' \mu}$ is evaluated
for scattering between all Fermi points.  The dressed phonon
frequencies $\Omega_{q\mu}$ are computed by using the
frozen-phonon approximation given in Eq.~(\ref{Eq:FPA}) by 
the displacement vectors from the dynamical matrix of graphene
given in Ref.~\onlinecite{Jishi93}.  The undressed frequencies
$\Omega_{q \mu}^{0}$, which enter the Fr\"{o}lich Hamiltonian in
Eq.~(\ref{frolich}), are then extracted by using the previously
computed quantities of the band structure and the electron-phonon
coupling, and the RPA analysis of the Peierls instability. This
method is elaborated in Sec.~\ref{sec:frequencies}.   After the
calculation of these quantities, the effective Fr\"{o}lich
Hamiltonian has been fully constructed.  The remarkable agreement
of the coefficients of the logarithmic divergences computed by
using quantities from the band structure and the electron-phonon
coupling with the frozen-phonon frequencies is a consistency check
for this method.

With the Fr\"{o}lich Hamiltonian, we then used the RPA analysis of
the Peierls instability (in Sec.~\ref{sec:CDW} ) and the McMillan
equation (in Sec.~\ref{sec:SC} and Appendix
\ref{Appendix:Migdal-Eliashberg}) to obtain the charge-density
wave and superconducting transition temperatures, the result with
the higher transition temperature being the dominant phase at low
temperatures.  For instance, when the CDW is dominant, the Fermi
surface will be destroyed around $T_{\rm CDW}$ eliminating
superconductivity altogether.  By this method, we provided an
exhaustive analysis of three types of CNTs: (5,0), (6,0), and
(5,5). The more conventional larger-radius (5,5) CNT was seen to
be stable against the CDW and SC transitions down to very low
temperatures ($\ll 1$K) if we only include electron-phonon
interactions. For the ultrasmall radius (5,0) and (6,0)
CNTs, however, the CDW was found to be the dominant phase, with
transition temperatures of $160$ and $6$ K respectively.  For both
of these CNTs, $2k_{F}$ is incommensurate with the underlying
lattice.  Furthermore, in contrast to larger radius CNTs which
have dominant electron-phonon coupling to the in-plane phonon
modes, the ultrasmall (5,0) and (6,0) CNTs were found to have
dominant coupling to the out-of-plane phonon modes (see
Fig.~\ref{fig:modes}), as seen from the direct computation of the
electron-phonon matrix elements $M_{k\tau k' \tau' \mu}$. This is
further supported by the frozen-phonon computation of frequencies
which show the most robust Kohn anomalies for these modes (see
Fig.~\ref{fig:modes}).

When we include the Coulomb interaction, for the (5,0) CNT we
find that the CDW order is suppressed much more strongly
than superconductivity.
More specifically, our analysis presented in Sec.~\ref{sec:coulomb}
 shows that the
CDW transition is pushed down to unobservably low
temperatures, whereas the superconducting $T_{\rm SC}$
is reduced to 1 K.  Hence our calculation supports the
possibility of observing superconductivity in ultrasmall CNTs.
It is quite foreseeable that a more
detailed model for the Coulomb interaction could raise $T_{\rm SC}$
to the value seen experimentally, especially considering the
exponential dependence of the superconducting transition temperature
on the Coulomb interaction strength.  For the (6,0) CNT,
we found that the CDW remains dominant when the Coulomb interactions
are included due to the weak Coulomb interaction between
electrons at the Fermi points, and occurs at around $T_{\rm CDW}=$5K.

\appendix

\section{The electron-phonon coupling vertices}
\label{Appendix:Mkk'}

The electron-phonon coupling matrix is given by
\begin{equation}
M_{k\tau k'\tau'\mu}
=N \bra{\psi_{k\tau}}
\sum_{i}\frac{\partial V}{\partial {\bf R}_{0i}} \cdot
\hat{\epsilon}_{q\mu}(i)\ket{\psi_{k'\tau'}}.
\end{equation}
One can see that the above expression can be evaluated by
using the finite difference formula
\begin{equation}
M_{k\tau k'\tau' \mu}=\frac{1}{u}\bra{\psi_{k\tau}}
(V^{q\mu}-V_{0})\ket{\psi_{k'\tau'}}.
\end{equation}
A method for calculating this expression with a plane-wave basis set
was previously developed.\cite{Lam86} This section will be
 devoted to describing how to calculate $M_{k\tau k'\tau' \mu}$ with a
tight-binding method.  We introduce the standard tight-binding notation
\begin{equation}
\ket{\psi_{k\tau}}=\sum_{il}A_{k\tau il}\ket{\chi_{kil}}
\end{equation}
\begin{equation}
\ket{\chi_{kil}}=\frac{1}{\sqrt{N}}\sum_{n}e^{ik\cdot R_{n}}\ket{\phi_{nil}}.
\end{equation}
Here $n$ runs over unit cells and $i$ runs over basis vectors in the unit cell
and $l$ over orbital type.  Because the kinetic energy operator will
be the same in the perturbed and unperturbed Hamiltonians, we can write
\begin{equation}
\label{Mkk'1}
M_{k\tau k'\tau' \mu}=\frac{1}{u}\bra{\psi_{k\tau}}(H^{q\mu}-\varepsilon_{F})\ket{\psi_{k'\tau'}}.
\end{equation}
The reason why we keep the $\varepsilon_{F}$ term which clearly is zero through orthogonality will become clear below.  Expanding the wave functions in the tight-binding basis set, we obtain
\begin{equation}
M_{k \tau k' \tau' \mu}=\frac{1}{u}
\sum_{ili'l'}A^{*}_{k\tau il}\bra{\chi_{kil}}
(H^{q\mu}-\varepsilon_{F})\ket{\chi_{k'i'l'}}A_{k'\tau' i'l'}.
\end{equation}
Now, we write 
$\ket{\chi_{kil}^{q\mu}}=\ket{\chi_{kil}}+\ket{\delta \chi_{kil}^{q \mu}}$ 
where the orbitals of $\ket{\chi_{kil}^{q\mu}}$ are centered on the perturbed lattice.  Inserting this into the above equation, we obtain
\begin{eqnarray}
M_{k \tau k' \tau' \mu}&=&\frac{1}{u}\sum_{ili'l'}
A^{*}_{k\tau il}\left (\bra{\chi_{kil}^{q\mu}}(H^{q\mu}-
\varepsilon_{F})\ket{\chi_{k'i'l'}^{q\mu}} \right.
\\ \nonumber
&-&
\left.  \left( \bra{\delta \chi_{kil}^{q\mu}}(H^{q\mu}-\varepsilon_{F})\ket{\chi_{k'i'l'}}
+{\rm h.c.} \right)\right)A_{k'\tau' i'l'}.
\end{eqnarray}
In the second term we can do the substitution
$H^{q\mu}\rightarrow H_{0}$ because the effect of doing this will be second order in $u$ and we are interested in an expression that is accurate to first order.  Then, this term will be
\begin{eqnarray}
&&\sum_{ili'l'}A^{*}_{k\tau il} \left( \bra{\delta \chi_{kil}}(H-\varepsilon_{F})\ket{\chi_{k'i'l'}}+h.c.\right)A_{k'\tau' i'l'}\nonumber
\\
&=&\sum_{il}A^{*}_{k\tau il}\bra{\delta \chi_{kil}}
(H-\varepsilon_{F}) \ket{\psi_{k'\tau'}}+h.c.=0.
\end{eqnarray}
So we finally have the expression
\begin{equation}
M_{k\tau k'\tau' \mu}=\frac{1}{u}\sum_{ili'l'}A^{*}_{k\tau il}
\bra{\chi_{kil}^{q\mu}}(H^{q\mu}-\varepsilon_{F})
\ket{\chi_{k'i'l'}^{q\mu}}A_{k'\tau'i'l'}.
\end{equation}
This expression can be computed by evaluating the tight-binding
Hamiltonian and overlap matrices
for the distorted lattice, evaluating the coefficients
$A_{k\tau il}$ and $A_{k'\tau' i'l'}$ of the wave functions for the
undistorted lattice, and performing the above sum.
There is a slight technical problem with the above method because
$k$ and $k'$ are not the same in the tight-binding matrix.  However,
it can be shown that the correct result will
be obtained by using
$\bra{\chi_{kil}^{q\mu}}H^{q\mu}\ket{\chi_{ki'l'}^{q\mu}}$ and
$\langle\chi_{kil}^{q\mu}\ket{\chi_{ki'l'}^{q\mu}}$ for the tight-binding and
overlap matrices  (or the similar expression with $k \rightarrow k'$)
in the limit of a large
supercell.  That is, when the distance over which neighboring atoms interact
is small compared to the length of a unit cell, this method becomes exact.
When we apply this method, we checked for convergence of the coupling as a
function of the unit cell size.

\section{Isotropic Eliashberg equations in 1D}
\label{Appendix:Migdal-Eliashberg}

Obtaining quantitative parameters of superconductors
described by the BCS theory like the transition temperature
and the wave vector-dependent superconducting gap from
microscopic models has developed into a powerful tool for
understanding experimentally realized systems as well as
even predicting new superconductors.\cite{Cohen86}  Though
excellent review articles exist, \cite{Allen82, Rainer86}
we will establish the key results of the theory below
in attempt to be as self-contained as possible.  We will
also show how to incorporate the electron-phonon coupling
into the phonon parameters which become important in 1d due
to the CDW instability.

\begin{figure}
\includegraphics[width=2.5in]{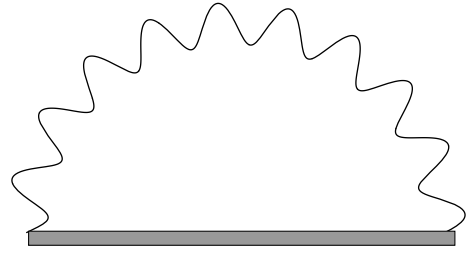}
\caption{Migdal's expression for the electronic self-energy.  
The thick line denotes the dressed electronic Green's function
and the wavy line denotes the phonon Green's functions.}
\label{fig:migdal}
\end{figure}

In the following to simplify notation, we will consider a single
band system only.
The central ingredient which, in principle, allows one to calculate
the superconducting transition temperature to high accuracy
is Migdal's theorem \cite{Migdal58} which allows one to evaluate
the electron self-energy with small error as
\begin{eqnarray}
\nonumber
{\bf \Sigma} (k,i\omega_{n})=-\frac{1}{\beta}&\sum_{k'n'\mu}&\tau_{3}
{\bf G} (k',i\omega_{n'})\tau_{3}|g_{kk'\mu}|^{2} 
\\
&\times&D_{0 \mu}(k-k',n-n').
\label{se}
\end{eqnarray}
This expression for the self-energy is shown in Fig.~\ref{fig:migdal}.
In this equation, $\beta$ is inverse temperature, $\tau_{i}$ are Pauli matrices
($i=0$ gives the identity matrix while $i=1,2,3$ give the $x,y,z$ Pauli matrices respectively), $D_{0}$ is the noninteracting phonon Green's function, and $\omega_{n}=\pi(2n+1)/\beta$ are the fermionic Matsubara frequencies.  The electronic Nambu-Green's function, a $2\times2$ matrix, is given by $\bf G\mit (k,i\omega_{n})=(i\omega_{n}\tau_{0}-\varepsilon_{k}\tau_{3}-\bf \Sigma \mit)^{-1}$.

 Now, we can expand $\bf \Sigma \mit$ in terms of Pauli matrices
\begin{equation}
{\bf \Sigma}=(1-Z)i\omega_{n}\tau_{0}+\phi \tau_{1}.
\end{equation}
We did not include the $\tau_3$ term because this just shifts the quasiparticle energies and similarly we neglected the $\tau_2$ term which can be eliminated
by a proper choice of phase for $\phi$.
Written in terms of these parameters, the Green's function becomes
\begin{equation}
{\bf G}=-\frac{Zi\omega_{n}\tau_{0}+\varepsilon_{k}\tau_{3}+\phi\tau_{1}}{(Z\omega_{n})^{2}+\varepsilon_{k}^{2}+\phi^{2}}.
\end{equation}
Inserting this into~\ref{se} we obtain
\begin{eqnarray}
{\bf \Sigma}&=&\frac{1}{\beta}\sum_{k'n'\mu}
\frac{Z'i\omega_{n'}\tau_{0}+\varepsilon_{k'}\tau_{3}-\phi'\tau_{1}}{(Z'\omega_{n'})^{2}+\varepsilon_{k'}^{2}+\phi'^{2}}
\nonumber
\\
&\times&|g_{kk'\mu}|^{2}D_{0\mu}(k-k',n-n').
\end{eqnarray}
Now, we insert the identity
$\int d\varepsilon \delta(\varepsilon-\varepsilon_k')$ into the above
expression to obtain
\begin{eqnarray}
{\bf \Sigma}&=&\frac{1}{\beta}\int d\varepsilon \sum_{k'n'\mu}\delta(\varepsilon-\varepsilon_k)
\frac{Z'i\omega_{n'}\tau_{0}+\varepsilon\tau_{3}-\phi'\tau_{1}}{(Z'\omega_{n'})^{2}+\varepsilon^{2}+\phi'^{2}}
\nonumber
\\
&\times&|g_{kk'\mu}|^{2}D_{0 \mu}(k-k',n-n').
\end{eqnarray}
The Lorentzian term in the integrand peaks very strongly at $\varepsilon=\varepsilon_{F}=0$ with width on the order of temperature.  Assuming that the rest of the integrand doesn't vary as rapidly about $\varepsilon=0$, we can replace $\delta(\varepsilon-\varepsilon_{k'})$ with $\delta(\varepsilon_{k'})$ and perform the $\varepsilon$ integral to obtain
\begin{equation}
\label{se1}
{\bf \Sigma} =\frac{\pi}{\beta}\sum_{k'n'\mu}\delta(\varepsilon_{k'})
\frac{Z'i\omega_{n'}\tau_{0}-\phi'\tau_{1}}{\sqrt{(Z'\omega_{n'})^{2}
+\phi'^{2}}}|g_{kk' \mu}|^{2}
D_{0 \mu}(k-k',n-n').
\end{equation}
This approximation can be seen to break down for small momentum (forward) scattering due to acoustic phonons.  This case will be discussed in
Appendix \ref{Appendix:Acoustic}.
When close to $T_{\rm SC}$, $\phi'$ will be small and can be neglected in the denominator of \ref{se1}.

Now we perform the so-called isotropic approximation.  Multiply both sides of~\ref{se1} by $\delta(\varepsilon_{k})/\nu_{\sigma}(0)$ where $\nu_{\sigma}(0)$
is the density of states at the Fermi level per spin and sum over $k$.
In the right-hand side of~\ref{se1} we then replace $Z(k',n')$
and $\phi(k',n')$ with their Fermi-surface averages $Z_{n'}$ and $\phi_{n'}$.
 This approximation is valid when the Fermi-surface is fairly isotropic.
Now by equating the coefficients of the matrices $\tau_{0}$ and $\tau_{1}$
we finally arrive at the equations
\begin{equation}
\label{eli1a}
Z_{n}=1+f_{n}s_{n}\sum_{n'}\lambda(n-n')s_{n'}
\end{equation}
\begin{equation}
\label{eli2a}
Z_{n}\Delta_{n}=\sum_{n'}(\lambda(n-n')-\mu_{\rm SC}^{*})f_{n'}\Delta_{n'}
\end{equation}
where $f_{n}=1/|2n+1|$, $s_{n}={\rm sgn}(2n+1 )$,
$\Delta_{n}=\phi_{n}/Z_{n}$, and
\begin{equation}
\label{lambda}
\lambda(n-n')=-\frac{1}{\nu_{\sigma}(0)}\sum_{kk' \mu} \delta(\varepsilon_{k})\delta(\varepsilon_{k'})|g_{kk' \mu}|^{2}D_{0 \mu}(k-k',n-n').
\end{equation}
The Coulomb pseudopotential $\mu_{\rm SC}^{*}$ was inserted to
 account for the bare electron-electron interaction that is not
included in our original
Hamiltonian Eq.~(\ref{frolich}).
The superconducting transition temperature  $T_{\rm SC}$
is the temperature at which nontrivial solutions for the gap $\Delta_{n}$
begin to appear.  Equations (\ref{eli1a}), (\ref{eli2a}), and (\ref{lambda})
 are known as the isotropic Eliashberg equations.\cite{Eliashberg60}
Input parameters have been calculated and the Eliashberg equations have been
solved to calculate $T_{\rm SC}$ for a variety of superconductors described
by the BCS theory.  We also note that a generalization to the case where
the Fermi surface is anisotropic is straightforward.\cite{Allen76}

Now, for typical three-dimensional solids the phonon frequencies
are affected very little by the electron-phonon coupling.  Therefore,
the above formalism where we have used the non-interacting
phonon Green's function $D_{0 \mu}$ works remarkably well in 3d.
This is not the case, however, in 1d where one is encountered with the
CDW instability.  A more accurate phonon Green's function is
given by
\begin{equation}
\label{dressed}
D_{\mu}(k,n)=\frac{2\Omega_{q\mu}^0}{(i\nu_n)^{2}-(\Omega_{q\mu})^{2}}
\end{equation}
where $\Omega_{q \mu}^0$ is the undressed frequency (without electron-phonon
coupling) and $\Omega_{q\mu}$ is the dressed frequency (which, as seen above
can have strong temperature dependence).  Calculating
the dressed phonon Green's function
can be challenging because one needs both  $\Omega_{q\mu}^0$ and
$\Omega_{q\mu}$.
However, we notice that when we substitute Eq.~(\ref{dressed})
into Eq.~(\ref{lambda}) we have the fortuitous
cancellation of $\Omega_{q\mu}^0$ in the numerator of $D_{\mu}(k,n)$ with
that in the denominator of $|g_{kk'\mu}|^{2}$.  Thus one sees that knowledge
of the undressed frequencies (which are significantly more difficult
to obtain) will not be necessary to construct
the Eliashberg equations.  By doing the substitution $\Omega_q^0
\rightarrow \Omega_{q\mu}$ in
Eqns.~(\ref{eli1a}), (\ref{eli2a}), and (\ref{lambda}),
one can thereby construct the ``dressed'' Eliashberg equations which
takes into account the influence of the electron-phonon coupling
on the phonon frequencies which is important in 1d.

Note also that since some modes will have temperature dependence,
the Eliashberg equations must be solved self consistently.  That
is, we must find a temperature such that the SC transition temperature
determined from the Eliashberg equations is the same as the
temperature used for the input dressed phonon frequencies.  This
can be done by iteration.  Furthermore, this method allows us to tell which
will be the dominant phase at low temperature of our system.
If we find a self-consistent solution of the Eliashberg equations
and $T_{\rm SC} > T_{\rm CDW}$, then superconductivity will be the dominant
correlation.  Otherwise, the system will prefer the CDW state.

Finally, we will write down an expression which approximately
solves the Eliashberg equations, originally developed by
McMillan
\begin{equation}
T_{\rm SC}=\frac{\langle \Omega \rangle}{1.20}
\exp \left[- \frac{ 1.04(1+\lambda_{\rm SC})}
{\lambda_{\rm SC}-\mu_{\rm SC}^{*}(1+0.62 \lambda_{\rm SC})} \right]
\end{equation}
where $\lambda_{\rm SC} \equiv \lambda (0)$.  From the above
analysis, we see that to be self-consistent, one should use
the dressed frequencies to evaluate $\lambda_{\rm SC}$.

\section{Incorporating $q\approx 0$ scattering from
acoustic phonons in the Eliashberg equations.}
\label{Appendix:Acoustic}

In this Appendix, we discuss in detail the role of acoustic
phonons for small-radius nanotubes.  Earlier theoretical analysis of
the electron-phonon interactions in 1d systems suggested that
acoustic phonons can play a dominant role
in stabilizing the superconducting state.\cite{Loss94, DeMartino03} 
We will
show, however, that since the dominant coupling comes
from optical modes, that this effect is not important
for the CNTs we study.

We now consider explicitly the 
contributions to $q \approx 0$ scattering processes
coupled to acoustic phonon modes 
which are not accounted for in
the approximations leading to \ref{se1}.
For the electron-phonon coupling to acoustic modes, we take 
\begin{equation}
|g_{q\mu}|^{2}= \frac{\gamma |q|/L}{1+(q/q_{0})^2}
\end{equation}
where $q_{0}$ is a cut-off of order $k_{F}$
and $L$ is the system length.  We also take $\Omega_{q\mu}=c|q|$
and $\varepsilon_k=v_F(|k|-k_F)$.  Inserting these quantities into \ref{se},
setting $Z=1$ for simplicity,
we obtain for the off-diagonal element
\begin{eqnarray}
\Delta_{n}^{(q \approx 0)}&=& \frac {1}{\beta}   
\sum_{n'}\frac{1}{2 \pi}\int dq \frac{\Delta_{n'}}{\omega_{n'}^2
+(v_{F}q)^{2}}
\\
&\times&
\frac{\gamma |q|}{1+(q/q_{0})^2} 
\cdot
\frac{2c|q|}{(\omega_n-\omega_{n'})^{2}+(c|q|)^2} \nonumber
\end{eqnarray}
This integral can be evaluated to give
\begin{eqnarray}
\Delta_{n}^{(q \approx 0)}&=&\frac{\gamma}{v_{F}\beta}
\sum_{n'}\frac{\Delta_{n'}}{|\omega_{n'}|c+|\omega_{n}-\omega_{n'}|v_F}
\\
&\times& \frac{q_{0}}{|\omega_{n}-\omega_{n'}|/c+q_{0}}
\cdot \frac{q_{0}}{|\omega_{n}-\omega_{n'}|/v_{F}+q_{0}}.
\nonumber
\end{eqnarray}
One then sees that scattering from $q \approx 0$ acoustic phonons 
gives an approximate contribution to
$\lambda_{\rm SC}$ (when $n=n'$) of 
$\lambda_{\rm SC}^{(q \approx 0)}={\gamma}/(\pi v_{F} c)$.

Now we consider the $q\approx 2k_{F}$ scattering process from the
same acoustic phonon.  For this process we obtain
\begin{eqnarray}
\Delta_{n}^{(q \approx 2k_{F})}\approx \frac {1}{\beta}
\sum_{n'} |g_{q=2k_{F},\mu}|^{2}D(2k_{F},n-n') \nonumber
\\
\times \frac{L}{2 \pi}\int dq \frac{\Delta_{n'}}{\omega_{n'}^2+(v_{F}q)^{2}}.
\end{eqnarray}
This integral can be evaluated to give
\begin{eqnarray}
\Delta_{n}^{(q \approx 2k_{F})}=\frac{4\gamma c k_{F}^{2}}{\beta  v_{F}} \sum_{n'}
\frac{1}{(\omega_{n}-\omega_{n'})^2+(c2k_{F})^{2}}\frac{1}{|\omega_{n'}|}.
\end{eqnarray}
One then finds that this gives a contribution of
$\lambda_{\rm SC}^{(q \approx 2k_{F})}=\gamma/(\pi v_{F} c)$ 
to $\lambda_{\rm SC}$ which
is exactly the same as the $q \approx 0$ scattering contribution.
Thus one sees that $q \approx 0$ 
scattering from acoustic phonons can
be very important in one-dimensional electron-phonon systems.  From
such a process the so-called Wentzel-Bardeen 
instability \cite{Wentzel51, Bardeen51,Engelsberg64}
can occur which has recently been studied in the 
context of CNTs.\cite{DeMartino03}  We also note
that a  similar analysis can be carried out for the 
optical phonons, and it is found
that the $q \approx 0$ processes are much smaller
than the $q \approx 2k_{F}$ process.

With the above method, we now see how to
include the contribution from $q \approx 0$
scattering into $\lambda_{\rm SC}$.  To do this, 
we simply double the contributions to $\lambda_{\rm SC}$
from $2k_{F}$ processes which couple to acoustic phonons
to include the $q \approx 0$ contribution.
In practice, we find that using this procedure actually
changes $\lambda_{\rm SC}$ by only a small amount.  
For instance, for the (5,0) CNT, $\lambda_{\rm SC}$ only
increases by less than 1\%.  This is because the dominant
contributions to $\lambda_{\rm SC}$ are
from coupling to the optical modes as discussed 
in Sec.~\ref{sec:results}.  

We also point out that presence of the Wentzel-Bardeen singularity would
significantly  renormalize the acoustic phonon mode frequencies of the
CNTs.  The fact that the calculated phonon frequencies using
the frozen-phonon approximation for the CNTs are quantitatively
similar to the analogous modes of graphene
as shown in Figs.~\ref{fig:pho5,0}, \ref{fig:pho6,0}, and \ref{fig:pho5,5}
further supports 
the the notion that the Wentzel-Bardeen instability is unimportant
in these systems.

\section{Limitations of non self-consistent method}
\label{Appendix:SCC}
In this Appendix, we will discuss the limitations of using a method 
in which the charge density is not evaluated  self-consistently.  
For simplicity, we will neglect the contribution from the exchange-correlation
energy $E_{\rm XC}$ in the Kohn Sham energy functional.

First we will consider the case of the equilibrium lattice structure.
For this, the self-consistent total energy is given by
\begin{eqnarray}
\label{Eq:KS}
E_{\rm SC}^{\rm eq}&=&\sum_{i}\bra{\psi_{i}}\left(\frac{{\bf p}^{2}}{2m}
+V_{\rm{ion}}^{\rm eq}({\bf r})
+\frac{1}{2}\int d^{3} r' \frac{n({\bf r}')}{|{\bf r}-{\bf r}'|}\right)\ket{\psi_{i}}
\nonumber
\\
&+&E_{\rm{ion-ion}}^{\rm eq}
\end{eqnarray}
where the charge-density is given by
$n({\bf r})=\sum_{i} |\psi_{i}({\bf r})|^{2}$, $V_{\rm ion}^{\rm eq}$ is the
ionic potential, and $E_{\rm ion-ion}^{\rm eq}$ is the ion-ion interaction.
In the above and in what follows, the $i$ summation is 
carried out only over occupied electronic states.  Applying the variational
principle to Eq.~\ref{Eq:KS} gives the equation for the wave functions $\ket{\psi_{i}}$
and therefore the charge-density $n({\bf r})$
\begin{equation}
{\cal H}^{\rm eq}[n]\ket{\psi_{i}}=\varepsilon_{i}\ket{\psi_{i}}
\end{equation}
where
\begin{equation}
{\cal H}^{\rm eq}[n]=\frac{{\bf p}^{2}}{2m}+V_{\rm{ion}}({\bf r})
+\int d^{3} r' \frac{n({\bf r}')}{|{\bf r}-{\bf r}'|}.
\end{equation}
In solving this equation, the charge density $n({\bf r})$
entering ${\cal H}^{\rm eq}[n]$ must be determined self-consistently
to agree with the eigenfunctions $\psi_{i}$.  Using this, the
self-consistent totally energy for the
equilibrium lattice is determined to be
\begin{equation}
E_{\rm SC}^{\rm eq}=\sum_{i} \bra{\psi_{i}}{\cal H}^{\rm eq}[n]\ket{\psi_{i}} +F^{\rm eq}[n]
\end{equation}
where 
\begin{equation}
F^{\rm eq}[n]=-\frac{1}{2}\int d^{3}r d^{3} r' \frac{n({\bf r}) n({\bf r}')}{|{\bf r}-{\bf r}'|}
+E_{\rm ion-ion}^{\rm eq}. 
\end{equation}
to be essentially the same as for non-interacting atoms.
In the tight-binding limit we expect the equilibrium electron density
to be essentially the same as for non-interacting atoms.  
If we denote the latter as $n_{0}({\bf r})$, we can replace
$n({\bf r})$ by $n_{0}({\bf r})$ in $E_{\rm SC}^{\rm eq}$ 
and expect the resulting non-self-consistent total energy
$E_{\rm NSC}^{\rm eq}$  to be quite 
close to the self-consistent total energy for the equilibrium lattice
structure:
\begin{equation}
\label{Eq:Eeq}
E_{\rm NSC}^{\rm eq} \approx E_{\rm SC}^{\rm eq}. 
\end{equation}
This approach is the basis for using an effective tight-binding model
to calculate band structures.

Such a method, however, breaks down when we consider a lattice
perturbed by a phonon.  In the presence of a lattice distortion,
the ionic potential changes to 
$V_{\rm ion}^{\rm dist}=V_{\rm ion}^{\rm eq}+\delta V_{\rm ion}$
which, in turn, makes the charge-density non-uniform
$n=n_{0}+\delta n$.  The energy of the distorted structure is
then
\begin{eqnarray}
E_{\rm SC}^{\rm dist}
&=&
\sum_{i}\bra{\psi_{i}}\left(\frac{{\bf p}^{2}}{2m}
+V_{\rm{ion}}^{\rm dist}({\bf r})
+\frac{1}{2}\int d^{3} r' \frac{n({\bf r}')}{|{\bf r}-{\bf r}'|}\right)\ket{\psi_{i}}
\nonumber
\\
&+&E_{\rm{ion-ion}}^{\rm dist}.
\end{eqnarray}
Now replacing $n$ with $n_{0}+\delta n$, this can be written
as
\begin{eqnarray}
\label{Eq:KS2}
E_{\rm SC}^{\rm dist}&=&\sum_{i}\bra{\psi_{i}}{\cal H}^{\rm dist}[n_{0}]\ket{\psi_{i}}
+F^{\rm dist}[n_{0}]
\\
\nonumber
&+&\frac{1}{2}\int d^{3} r d^{3} r' \left(  \frac{1}{|{\bf r}-{\bf r}'|}
\right) \delta n({\bf r}) \delta n({\bf r'})
\end{eqnarray}
where ${\cal H}^{\rm dist}$ and $F^{\rm dist}$ are given by
${\cal H}^{\rm eq}$ and $F^{\rm eq}$ defined above with $V_{\rm ion}^{\rm eq}$
and $E_{\rm ion-ion}^{\rm eq}$ replaced by $V_{\rm ion}^{\rm dist}$
and $E_{\rm ion-ion}^{\rm dist}$.  The first two terms on the right
of Eq.~\ref{Eq:KS2} can be seen to be the total energy 
of the distorted structure computed with the non-self-consistent method.  
We therefore obtain
\begin{equation}
E_{\rm SC}^{\rm dist}=E_{\rm NSC}^{\rm dist}+
\frac{1}{2}\int d^{3} r d^{3} r' \left(  \frac{1}{|{\bf r}-{\bf r}'|}
\right) \delta n({\bf r}) \delta n({\bf r'}).
\end{equation}
Subtracting Eq.~\ref{Eq:Eeq} from this then gives
\begin{equation}
\Delta E_{\rm SC}=\Delta E_{\rm NSC}+
\frac{1}{2}\int d^{3} r d^{3} r' \left(  \frac{1}{|{\bf r}-{\bf r}'|}
\right) \delta n({\bf r}) \delta n({\bf r'})
\end{equation}
where 
$\Delta E_{\rm SC,NSC}=E_{\rm SC,NSC}^{\rm dist}-E_{\rm SC,NSC}^{\rm eq}$.
Rewriting the second term in momentum space gives
\begin{equation}
\label{Eq:Hartree_drop}
\Delta E_{\rm SC}=\Delta E_{\rm NSC}+
\frac{1}{2}\int \frac{d^{3}q}{(2 \pi)^3} V(q)
 |\delta n_{\bf q}|^{2}.
\end{equation}
which then makes it clear that
$\Delta E_{\rm SC}>\Delta E_{\rm NSC}$.  
So we see that using the a non-self-consistent
method to calculate phonon frequencies by the frozen-phonon approximation
will underestimate the phonon frequencies. More specifically, 
in a non-self-consistent
method, the Hartree term displayed in Eq.~(\ref{Eq:Hartree_drop}) 
is not accounted for.
This should be particularly
important in the vicinity of a CDW instability, where there will
be a larger response of the charge distribution to a lattice distortion.

\section{Derivation of Eq.~(\ref{Eq:BSCoulomb})}
\label{Appendix:CoulDerivation}
In this Appendix, we will derive Eq.~(\ref{Eq:BSCoulomb}) by
evaluating the integral appearing in Eq.~(\ref{Eq:Coulombintegral}).
To estimate this Coulomb interaction integral, we will take the
tight-binding wave function of graphene
\begin{equation}
\label{Eq:Graphitewf}
\psi_{{\bf k}\gamma}({\bf r})=\frac{1}{\sqrt{N}}\sum_{n} e^{i {\bf k}\cdot
{\bf R}_{n}}\frac{1}{\sqrt{2}}\left(\gamma \frac{f({\bf k})}{|f({\bf k})|}
\phi_{n1}({\bf r})+\phi_{n2}({\bf r})\right).
\end{equation}
Now ${\bf k}$ is a two dimensional vector in reciprocal space of the
graphene lattice and $\gamma= \pm 1$ corresponds to the conduction and valence
bands.
Orbitals centered
on the first and second carbon atoms respectively in the $n$th unit cell
are given by  $\phi_{n1}({\bf r})$ and $\phi_{n2}({\bf r})$ respectively,
and $f({\bf k})$ is given by
$f({\bf k})=1+e^{-i {\bf k}\cdot {\bf a}_{1}}+e^{-i {\bf k}\cdot {\bf a}_{2}}$
where ${\bf a}_1$ and ${\bf a}_2$ are the lattice vectors of graphene.  For
metallic large radius CNTs, the Fermi points correspond to
${\bf K}=\frac{1}{3}({\bf b}_1-{\bf b}_2)$ and
${\bf K}'=\frac{2}{3}({\bf b}_1-{\bf b}_2)$ where
${\bf b}_1$ and ${\bf b}_2$ are the reciprocal lattice vectors corresponding
to ${\bf a}_1$ and ${\bf a}_2$.  For these points, we have
$f({\bf K})=f({\bf K}')=0$.  However, for the smaller
radius CNTs we study, as indicated by the failure of the zone-folding method,
the Fermi points are shifted away from ${\bf K}$ and ${\bf K}'$.  We
denote the Fermi points of the inner band $\tau_{a}$ of the (5,0) CNT
by ${\bf k}_{\tau_{a}+}={\bf K}+k_{x}{\bf \hat{x}}-k_{y}{\bf \hat{y}}$ and
${\bf k}_{\tau_{a}-}={\bf K}-k_{x}{\bf \hat{x}}-k_{y}{\bf \hat{y}}$ and
for the other inner band $\tau_{b}$ by
${\bf k}_{\tau_{b}+}={\bf K}'+k_{x}{\bf \hat{x}}+k_{y}{\bf \hat{y}}$ and
${\bf k}_{\tau_{b}-}={\bf K}'-k_{x}{\bf \hat{x}}+k_{y}{\bf \hat{y}}$
where the $x$-direction is still along
the CNT axis and the $y$-direction is along the perimeter.

For backward scattering, we take
$q\approx 2k_{F}, k\approx-k_{F}, k'\approx k_{F}$.
Keeping only products of Carbon orbitals centered on the same
atom, we obtain
\begin{eqnarray}
\psi_{k+q \tau_{a}}^{*}({\bf r})\psi_{k \tau_{a}}({\bf r})
&\approx&
\frac{1}{N}\sum_{n}e^{-iq{\bf\hat{x}}\cdot {\bf R}_{n}}
\frac{1}{2}
\\
\nonumber
&\times& \left( \frac{f^{*}({\bf k}_{\tau_{a}+})f({\bf k}_{\tau_{a}-})}
{|f^{*}({\bf k}_{\tau_{a}+})f({\bf k}_{\tau_{a}-})|}
 |\phi_{n1}({\bf r})|^{2} \right.
\\
& &  +
\left. |\phi_{n2}({\bf r})|^2 \frac{^{}_{}}{^{}_{}} \right).
\nonumber
\end{eqnarray}
Now we make use of the slow variation of $e^{-iqx}$
compared to the localized orbitals to write
\begin{eqnarray}
\psi_{k+q \tau}^{*}({\bf r})\psi_{k \tau}({\bf r})
&\approx&
e^{-iqx}\frac{1}{N}\sum_{n}
\frac{1}{2}
\\
\nonumber
&\times& \left(
\frac{f^{*}({\bf k}_{\tau_{a}+})f({\bf k}_{\tau_{a}-})}
{|f^{*}({\bf k}_{\tau_{a}+})f({\bf k}_{\tau_{a}-})|}
|\phi_{n1}({\bf r})|^{2} \right.
\\
& &  + \left.   e^{iq{\bf\hat{x}}
\cdot {\bf t}}|\phi_{n2}({\bf r})|^2 \frac{^{}_{}}{^{}_{}}
\right)
\nonumber
\end{eqnarray}
where ${\bf t}=\frac{1}{\sqrt{3}}a{\bf \hat{x}}$
is the basis vector for the second Carbon
atom in the primitive unit cell.  Finally, in evaluating the integral
in Eq.~(\ref{Eq:Coulombintegral})
it is sufficient to replace the functions
$\frac{1}{N}\sum_{n} |\phi_{n1,2}({\bf r})|^{2}$
which vary more rapidly than $V({\bf r})$ by their average values.
That is, we substitute
\begin{eqnarray}
\psi_{k+q \tau}^{*}({\bf r})\psi_{k \tau}({\bf r}) &\rightarrow&
\\
\frac{1}{2\pi R L}e^{-iqx}&\frac{1}{2}&
\left(
\frac{f^{*}({\bf k}_{\tau_{a}+})f({\bf k}_{\tau_{a}-})}
{|f^{*}({\bf k}_{\tau_{a}+})f({\bf k}_{\tau_{a}-})|}
+e^{iq{\bf\hat{x}}\cdot {\bf t}}\right).
\nonumber
\end{eqnarray}
Using the same approximations for the factor
$\psi_{k'-q}({\bf r}')\psi_{k'}({\bf r}')$, we
obtain for the Coulomb interaction
\begin{eqnarray}
V_{q\tau_{a}\tau_{a}}
&\approx&
\frac{1}{4}\left|
\frac{f^{*}({\bf k}_{\tau_{a}+})f({\bf k}_{\tau_{a}-})}
{|f^{*}({\bf k}_{\tau_{a}+})f({\bf k}_{\tau_{a}-})|}
+e^{iq{\bf\hat{x}}\cdot {\bf t}}\right|^{2}
\\
&\times&\frac{1}{L^{2}}\int dx dx' e^{-iq(x-x')}
\nonumber
\\
&\times&
 \int_{0}^{2 \pi R}\frac{dy}{2\pi R}\int_{0}^{2 \pi R}\frac{dy'}{2\pi R}
V({\bf r}-{\bf r}').
\end{eqnarray}

We will now evaluate the prefactor in this equation for the
inner band of the (5,0) and (6,0) CNTs.
Using the calculated Fermi points along
with the zone-folding method, we obtain
$k_{x}=\pm \frac{0.11}{\sqrt{3}}\frac{2\pi}{a}$ and
$k_{y}=\frac{1}{15}\frac{2\pi}{a}$ for the (5,0) CNT.  From this
we obtain
\begin{equation}
\frac{1}{4}\left|
\frac{f^{*}({\bf k}_{\tau_{a}+})f({\bf k}_{\tau_{a}-})}
{|f^{*}({\bf k}_{\tau_{a}+})f({\bf k}_{\tau_{a}-})|}
+e^{iq{\bf\hat{x}}\cdot {\bf t}}\right|^{2}=0.59
\end{equation}
For the (6,0) CNT the Fermi points are 
$k_{x}=\pm \frac{0.076}{\sqrt{3}}\frac{2\pi}{a}$ and
$k_{y}=0$.
This gives
\begin{equation}
\frac{1}{4}\left|
\frac{f^{*}({\bf k}_{\tau_{a}+})f({\bf k}_{\tau_{a}-})}
{|f^{*}({\bf k}_{\tau_{a}+})f({\bf k}_{\tau_{a}-})|}
+e^{iq{\bf\hat{x}}\cdot {\bf t}}\right|^{2}=0.0016
\end{equation}
which is smaller due to the different symmetry of
the wavefunctions at the Fermi points.
These are the values of the prefactor $\gamma$ 
appearing in Eq.~(\ref{Eq:BSCoulomb}).


\begin{thebibliography}{10}
\providecommand*{\bibinfo}[2]{#2}
\providecommand*{\eprint}[1]{#1}
\providecommand*{\url}[1]{#1}
\bibitem{Ijima91}
\bibinfo{author}{S.~Ijima}, \bibinfo{journal}{Nature}
  \bibinfo{volume}{\textbf{54}}, \bibinfo{pages}{56} (\bibinfo{date}{1991}).
\bibitem{Egger00}
\bibinfo{author}{R.~Egger}, \bibinfo{author}{A.~Bachtold},
  \bibinfo{author}{M.~Fuhrer}, and \bibinfo{author}{M.~Bockrath}, in
  \bibinfo{editors}{R.~Haug and H.~Schoeller}, eds., \emph{Interacting
  Electrons in Nanostructures} (\bibinfo{publisher}{Springer Verlag},
  \bibinfo{year}{2001}).
\bibitem{Kociak01}
\bibinfo{author}{M.~Kociak}, \bibinfo{author}{A.~Y. Kasumov},
  \bibinfo{author}{S.~Gueron}, \bibinfo{author}{B.~Reulet},
  \bibinfo{author}{I.~I. Khodos}, \bibinfo{author}{Y.~B. Gorbatov},
  \bibinfo{author}{V.~T. Volkov}, \bibinfo{author}{L.~Vaccarini}, and
  \bibinfo{author}{H.~Bouchiat}, \bibinfo{journal}{Phys. Rev. Lett.}
  \bibinfo{volume}{\textbf{86}}, \bibinfo{pages}{2416} (\bibinfo{date}{2001}).
\bibitem{Tang01}
\bibinfo{author}{Z.~Tang}, \bibinfo{author}{L.~Zhang},
  \bibinfo{author}{N.~Wang}, \bibinfo{author}{X.~Zhang},
  \bibinfo{author}{G.~Wen}, \bibinfo{author}{G.~Li}, \bibinfo{author}{J.~Wang},
  \bibinfo{author}{C.~Chan}, and \bibinfo{author}{P.~Sheng},
  \bibinfo{journal}{Science} \bibinfo{volume}{\textbf{292}},
  \bibinfo{pages}{2462} (\bibinfo{date}{2001}).
\bibitem{Mintmire92}
\bibinfo{author}{J.~W. Mintmire}, \bibinfo{author}{B.~I. Dunlap}, and
  \bibinfo{author}{C.~T. White}, \bibinfo{journal}{Phys. Rev. Lett.}
  \bibinfo{volume}{\textbf{68}}, \bibinfo{pages}{631} (\bibinfo{date}{1992}).
\bibitem{Huang96}
\bibinfo{author}{Y.~Huang}, \bibinfo{author}{M.~Okada},
  \bibinfo{author}{K.~Tanaka}, and \bibinfo{author}{T.~Yamabe},
  \bibinfo{journal}{Solid State Commun.} \bibinfo{volume}{\textbf{97}},
  \bibinfo{pages}{303} (\bibinfo{date}{1996}).
\bibitem{Sedeki00}
\bibinfo{author}{A.~Sedeki}, \bibinfo{author}{L.~G. Caron}, and
  \bibinfo{author}{C.~Bourbonnais}, \bibinfo{journal}{Phys. Rev. B}
  \bibinfo{volume}{\textbf{62}}, \bibinfo{pages}{6975} (\bibinfo{date}{2000}).
\bibitem{Dubay02}
\bibinfo{author}{O.~Dubay}, \bibinfo{author}{G.~Kreese}, and
  \bibinfo{author}{H.~Kuzmany}, \bibinfo{journal}{Phys. Rev. Lett.}
  \bibinfo{volume}{\textbf{88}}, \bibinfo{pages}{235506}
  (\bibinfo{date}{2002}).
\bibitem{Benedict95}
\bibinfo{author}{L.~X. Benedict}, \bibinfo{author}{V.~H. Crespi},
  \bibinfo{author}{S.~G. Louie}, and \bibinfo{author}{M.~L. Cohen},
  \bibinfo{journal}{Phys. Rev. B} \bibinfo{volume}{\textbf{52}},
  \bibinfo{pages}{14935} (\bibinfo{date}{1995}).
\bibitem{Sedeki02}
\bibinfo{author}{A.~Sedeki}, \bibinfo{author}{L.~G. Caron}, and
  \bibinfo{author}{C.~Bourbonnais}, \bibinfo{journal}{Phys. Rev. B}
  \bibinfo{volume}{\textbf{65}}, \bibinfo{pages}{140515}
  (\bibinfo{date}{2002}).
\bibitem{Byczuk02}
\bibinfo{author}{K.~Byczuk} (\bibinfo{date}{2002}), \eprint{condmat/0206086}.
\bibitem{Gonzalez02}
\bibinfo{author}{J.~Gonzalez}, \bibinfo{journal}{Phys. Rev. Lett.}
  \bibinfo{volume}{\textbf{88}}, \bibinfo{pages}{76403} (\bibinfo{date}{2002}).
\bibitem{Kamide03}
\bibinfo{author}{K.~Kamide}, \bibinfo{author}{T.~Kimura},
  \bibinfo{author}{M.~Nishida}, and \bibinfo{author}{S.~Kurihara}
  (\bibinfo{date}{2003}), \eprint{condmat/0301115}.
\bibitem{Schrieffer64}
\bibinfo{author}{J.~R. Schrieffer}, \bibinfo{title}{\emph{Theory of
  Superconductivity}} (\bibinfo{publisher}{Perseus Books},
  \bibinfo{year}{1963}).
\bibitem{Saito98}
\bibinfo{author}{R.~Saito}, \bibinfo{author}{G.~Dresselhaus}, and
  \bibinfo{author}{M.~S. Dresselhaus}, \bibinfo{title}{\emph{Physical
  Properties of Carbon Nanotubes}} (\bibinfo{publisher}{Imperial College
  Press}, London, \bibinfo{year}{1998}).
\bibitem{Mehl96}
\bibinfo{author}{M.~J. Mehl} and \bibinfo{author}{D.~A. Papaconstantopoulos},
  \bibinfo{journal}{Phys. Rev. B} \bibinfo{volume}{\textbf{54}},
  \bibinfo{pages}{4519} (\bibinfo{date}{1996}).
\bibitem{Blase94}
\bibinfo{author}{X.~Blase}, \bibinfo{author}{L.~X. Benedict},
  \bibinfo{author}{E.~L. Shirley}, and \bibinfo{author}{S.~G. Louie},
  \bibinfo{journal}{Phys. Rev. Lett.} \bibinfo{volume}{\textbf{72}},
  \bibinfo{pages}{1878} (\bibinfo{date}{1994}).
\bibitem{Kohn59}
\bibinfo{author}{W.~Kohn}, \bibinfo{journal}{Phys. Rev. Lett.}
  \bibinfo{volume}{\textbf{2}}, \bibinfo{pages}{393} (\bibinfo{date}{1959}).
\bibitem{Peierls55}
\bibinfo{author}{R.~E. Peierls}, \bibinfo{title}{\emph{Quantum Theory of
  Solids}} (\bibinfo{publisher}{Oxford University}, \bibinfo{year}{1955}).
\bibitem{Levin74}
\bibinfo{author}{K.~Levin}, \bibinfo{author}{D.~L. Mills}, and
  \bibinfo{author}{S.~L. Cunningham}, \bibinfo{journal}{Phys. Rev. B}
  \bibinfo{volume}{\textbf{10}}, \bibinfo{pages}{3821} (\bibinfo{date}{1974}).
\bibitem{Egger98}
\bibinfo{author}{R.~Egger} and \bibinfo{author}{A.~O. Gogolin},
  \bibinfo{journal}{Eur. Phys. J. B} \bibinfo{volume}{\textbf{3}},
  \bibinfo{pages}{281} (\bibinfo{date}{1998}).
\bibitem{Hohenberg64}
\bibinfo{author}{P.~Hohenberg} and \bibinfo{author}{W.~Kohn},
  \bibinfo{journal}{Phys. Rev.} \bibinfo{volume}{\textbf{136}},
  \bibinfo{pages}{B864} (\bibinfo{date}{1964}).
\bibitem{Kohn65}
\bibinfo{author}{W.~Kohn} and \bibinfo{author}{L.~Sham},
  \bibinfo{journal}{Phys. Rev.} \bibinfo{volume}{\textbf{140}},
  \bibinfo{pages}{A1133} (\bibinfo{date}{1965}).
\bibitem{Lam86}
\bibinfo{author}{P.~Lam}, \bibinfo{author}{M.~Dacorogna}, and
  \bibinfo{author}{M.~Cohen}, \bibinfo{journal}{Phys. Rev. B}
  \bibinfo{volume}{\textbf{34}}, \bibinfo{pages}{5065} (\bibinfo{date}{1986}).
\bibitem{Jishi93}
\bibinfo{author}{R.~A. Jishi}, \bibinfo{author}{L.~Venkataraman},
  \bibinfo{author}{M.~S. Dresselhaus}, and \bibinfo{author}{G.~Dresselhaus},
  \bibinfo{journal}{Chem. Phys. Lett.} \bibinfo{volume}{\textbf{209}},
  \bibinfo{pages}{77} (\bibinfo{date}{1993}).
\bibitem{Yin82}
\bibinfo{author}{M.~T. Yin} and \bibinfo{author}{M.~L. Cohen},
  \bibinfo{journal}{Phys. Rev. B} \bibinfo{volume}{\textbf{26}},
  \bibinfo{pages}{5668} (\bibinfo{date}{1982}).
\bibitem{Peng00}
\bibinfo{author}{L.~M. Peng}, \bibinfo{author}{Z.~L. Zhang},
  \bibinfo{author}{Z.~Q. Xue}, \bibinfo{author}{Q.~D. Wu},
  \bibinfo{author}{Z.~N. Gu}, and \bibinfo{author}{D.~G. Pettifor},
  \bibinfo{journal}{Phys. Rev. Lett.} \bibinfo{volume}{\textbf{85}},
  \bibinfo{pages}{3249} (\bibinfo{date}{2000}).
\bibitem{Li01}
\bibinfo{author}{Z.~M. Li}, \bibinfo{author}{Z.~K. Tang},
  \bibinfo{author}{H.~J. Liu}, \bibinfo{author}{N.~Wang},
  \bibinfo{author}{C.~T. Chan}, \bibinfo{author}{R.~Saito},
  \bibinfo{author}{S.~Okada}, \bibinfo{author}{G.~D. Li},
  \bibinfo{author}{J.~S. Chen}, \bibinfo{author}{N.~Nagasawa}, \emph{et~al.},
  \bibinfo{journal}{Phys. Rev. Lett.} \bibinfo{volume}{\textbf{87}},
  \bibinfo{pages}{127401} (\bibinfo{date}{2001}).
\bibitem{Dresselhaus81}
\bibinfo{author}{M.~S. Dresselhaus} and \bibinfo{author}{G.~Dresselhaus},
  \bibinfo{journal}{Adv. Phys.} \bibinfo{volume}{\textbf{30}},
  \bibinfo{pages}{139} (\bibinfo{date}{1981}).
\bibitem{Caron00}
\bibinfo{author}{A.~Sedeki}, \bibinfo{author}{L.~G. Caron}, and
  \bibinfo{author}{C.~Bourbonnais}, \bibinfo{journal}{Phys. Rev. B}
  \bibinfo{volume}{\textbf{62}}, \bibinfo{pages}{6975} (\bibinfo{date}{2000}).
\bibitem{Heeger79}
\bibinfo{author}{A.~J. Heeger}, in \bibinfo{editors}{J.~T. Devreese, R.~P.
  Evrard, and V.~E. van Doren}, eds., \emph{Highly Conducting One-Dimensional
  Solids} (\bibinfo{publisher}{Plenum Press}, New York, \bibinfo{year}{1979}).
\bibitem{Ginzburg82}
\bibinfo{editors}{\bibinfo{author}{V.~L. Ginzburg} and \bibinfo{author}{D.~A.
  Kirzhnits}}, eds., \bibinfo{title}{\emph{High Temperature Superconductivity}}
  (\bibinfo{publisher}{The Consultants Bureau}, New York,
  \bibinfo{year}{1982}).
\bibitem{McMillan68}
\bibinfo{author}{W.~L. McMillan}, \bibinfo{journal}{Phys. Rev.}
  \bibinfo{volume}{\textbf{167}}, \bibinfo{pages}{1967} (\bibinfo{date}{1967}).
\bibitem{Allen82}
\bibinfo{author}{P.~B. Allen} and \bibinfo{author}{B.~Mitrovic}, in
  \bibinfo{editors}{F.~Seitz, D.~Turnbull, and H.~Ehrenreich}, eds.,
  \emph{Solid State Physics} (\bibinfo{publisher}{Academic Press}, New York,
  \bibinfo{year}{1982}), \bibinfo{volume}{vol.~37}.
\bibitem{Wentzel51}
\bibinfo{author}{G.~Wentzel}, \bibinfo{journal}{Phys. Rev.}
  \bibinfo{volume}{\textbf{83}}, \bibinfo{pages}{168} (\bibinfo{date}{1951}).
\bibitem{Bardeen51}
\bibinfo{author}{J.~Bardeen}, \bibinfo{journal}{Rev. Mod. Phys.}
  \bibinfo{volume}{\textbf{23}}, \bibinfo{pages}{261} (\bibinfo{date}{1951}).
\bibitem{Engelsberg64}
\bibinfo{author}{S.~Engelsberg} and \bibinfo{author}{B.~B. Varga},
  \bibinfo{journal}{Phys. Rev.} \bibinfo{volume}{\textbf{136}},
  \bibinfo{pages}{A1582} (\bibinfo{date}{1964}).
\bibitem{Benedict95_2}
\bibinfo{author}{L.~X. Benedict}, \bibinfo{author}{V.~H. Crespi},
  \bibinfo{author}{S.~G. Louie}, and \bibinfo{author}{M.~L. Cohen},
  \bibinfo{journal}{Phys. Rev. B} \bibinfo{volume}{\textbf{52}},
  \bibinfo{pages}{8541} (\bibinfo{date}{1995}).
\bibitem{Inoshita77}
\bibinfo{author}{T.~Inoshita}, \bibinfo{author}{K.~Nakao}, and
  \bibinfo{author}{H.~Kamimura}, \bibinfo{journal}{J. Phys. Soc. Jpn.}
  \bibinfo{volume}{\textbf{43}}, \bibinfo{pages}{1237} (\bibinfo{date}{1977}).
\bibitem{Antropov92}
\bibinfo{author}{V.~P. Antropov}, \bibinfo{author}{O.~Gunnarsson}, and
  \bibinfo{author}{O.~Jepsen}, \bibinfo{journal}{Phys. Rev. B}
  \bibinfo{volume}{\textbf{46}}, \bibinfo{pages}{13647} (\bibinfo{date}{1992}).
\bibitem{Voit87}
\bibinfo{author}{J.~Voit} and \bibinfo{author}{H.~J. Schulz},
  \bibinfo{journal}{Phys. Rev. B} \bibinfo{volume}{\textbf{36}},
  \bibinfo{pages}{968} (\bibinfo{date}{1987}).
\bibitem{Voit88}
\bibinfo{author}{J.~Voit} and \bibinfo{author}{H.~J. Schulz},
  \bibinfo{journal}{Phys. Rev. B} \bibinfo{volume}{\textbf{37}},
  \bibinfo{pages}{10068} (\bibinfo{date}{1988}).
\bibitem{Grest76}
\bibinfo{author}{G.~S. Grest}, \bibinfo{author}{E.~Abrahams},
  \bibinfo{author}{S.-T. Chui}, \bibinfo{author}{P.~A. Lee}, and
  \bibinfo{author}{A.~Zawadowski}, \bibinfo{journal}{Phys. Rev. B}
  \bibinfo{volume}{\textbf{14}}, \bibinfo{pages}{1225} (\bibinfo{date}{1976}).
\bibitem{Langer67}
\bibinfo{author}{J.~S. Langer} and \bibinfo{author}{V.~Ambegaokar},
  \bibinfo{journal}{Phys. Rev.} \bibinfo{volume}{\textbf{164}},
  \bibinfo{pages}{498} (\bibinfo{date}{1967}).
\bibitem{McCumber70}
\bibinfo{author}{D.~E. McCumber} and \bibinfo{author}{B.~I. Halperin},
  \bibinfo{journal}{Phys. Rev. B} \bibinfo{volume}{\textbf{1}},
  \bibinfo{pages}{1054} (\bibinfo{date}{1970}).
\bibitem{caron94}
\bibinfo{author}{L.~G. Caron}, in \bibinfo{editors}{J.-P. Farges}, ed.,
  \emph{Organic Conductors: Fundamentals and Applications}
  (\bibinfo{publisher}{Marcel Dekker}, New York, \bibinfo{year}{1994}).
\bibitem{Kasumov99}
\bibinfo{author}{A.~Kasumov}, \bibinfo{author}{R.~Deblock},
  \bibinfo{author}{M.~Kociak}, \bibinfo{author}{B.~Reulet},
  \bibinfo{author}{H.~Bouchiat}, \bibinfo{author}{I.~Khodos},
  \bibinfo{author}{Y.~Gorbatov}, \bibinfo{author}{V.~Volkov},
  \bibinfo{author}{C.~Journet}, , \emph{et~al.}, \bibinfo{journal}{Science}
  \bibinfo{volume}{\textbf{284}}, \bibinfo{pages}{1508} (\bibinfo{date}{1999}).
\bibitem{Morpurgo99}
\bibinfo{author}{A.~Morpurgo}, \bibinfo{author}{J.~Kong},
  \bibinfo{author}{C.~Marcus}, and \bibinfo{author}{H.~Dai},
  \bibinfo{journal}{Science} \bibinfo{volume}{\textbf{286}},
  \bibinfo{pages}{263} (\bibinfo{date}{1999}).
\bibitem{Hertel00}
\bibinfo{author}{T.~Hertel} and \bibinfo{author}{G.~Moos},
  \bibinfo{journal}{Phys. Rev. Lett.} \bibinfo{volume}{\textbf{84}},
  \bibinfo{pages}{5002} (\bibinfo{date}{2000}).
\bibitem{Cohen86}
\bibinfo{author}{M.~L. Cohen}, \bibinfo{journal}{Science}
  \bibinfo{volume}{\textbf{234}}, \bibinfo{pages}{549} (\bibinfo{date}{1986}).
\bibitem{Rainer86}
\bibinfo{author}{D.~Rainer}, \bibinfo{journal}{Prog. in Low Temp. Phys.}
  \bibinfo{volume}{\textbf{10}}, \bibinfo{pages}{371} (\bibinfo{date}{1986}).
\bibitem{Migdal58}
\bibinfo{author}{A.~B. Migdal}, \bibinfo{journal}{Zh. Eksp. Teor. Fiz. [Sov.
  Phys. JETP]} \bibinfo{volume}{\textbf{34}}, \bibinfo{pages}{1438}
  (\bibinfo{date}{1958}).
\bibitem{Eliashberg60}
\bibinfo{author}{G.~M. Eliashberg}, \bibinfo{journal}{Zh. Eksp. Teor. Fiz.
  [Sov. Phys. JETP]} \bibinfo{volume}{\textbf{11}}, \bibinfo{pages}{696}
  (\bibinfo{date}{1960}).
\bibitem{Allen76}
\bibinfo{author}{P.~B. Allen}, \bibinfo{journal}{Phys. Rev. B}
  \bibinfo{volume}{\textbf{13}}, \bibinfo{pages}{1416} (\bibinfo{date}{1976}).
\bibitem{Loss94}
\bibinfo{author}{D.~Loss} and \bibinfo{author}{T.~Martin},
  \bibinfo{journal}{Phys. Rev. B} \bibinfo{volume}{\textbf{50}},
  \bibinfo{pages}{12160} (\bibinfo{date}{1994}).
\bibitem{DeMartino03}
\bibinfo{author}{A.~DeMartino} and \bibinfo{author}{R.~Egger},
  \bibinfo{journal}{Phys. Rev. B} \bibinfo{volume}{\textbf{67}},
  \bibinfo{pages}{235418} (\bibinfo{date}{2003}).

\end{thebibliography}
\end{document}